\documentclass[journal]{IEEEtran}
\usepackage[normalem]{ulem}
\usepackage{bm}
\usepackage{physics}
\usepackage{subcaption}
\usepackage{graphicx}
\usepackage[table]{xcolor}
\usepackage{atbegshi}
\usepackage{amsfonts}
\usepackage{multirow}
\usepackage{hhline}  % added by Jarek

 at 12truept
\newcommand{\nn}{{\bf{n}}}

\newcommand{\vv}{{\bf v}}

\renewcommand{\AA}{{\mathbf G}}
\newcommand{\sh}{\hat{s}}
\newcommand{\BBB }{{\mathbf B}}  %John's  A

\newcommand{\MM}{{\mathcal M}}

\newcommand{\SSS}{\bf{S}}
\renewcommand{\Im}{\text{\rm Im}}

\newcommand{\bLambda}{{\bf \Lambda}}

\newcommand{{\pp}}{p}
\newcommand{{\ppp}}{{\bf p}}
\newcommand {\mytrace}{\mbox{{\rm Tr}}}
\newcommand{\df}{:=} 
\newcommand{\beq}{\begin{equation}} 
\newcommand{\eeq}{\end{equation}}

\AtBeginDocument{\AtBeginShipoutNext{\AtBeginShipoutDiscard}}
\begin{document}

%
% paper title
% Titles are generally capitalized except for words such as a, an, and, as,
% at, but, by, for, in, nor, of, on, or, the, to and up, which are usually
% not capitalized unless they are the first or last word of the title.
% Linebreaks \\ can be used within to get better formatting as desired.
% Do not put math or special symbols in the title.
\title{Optimal launch states for the measurement of principal modes in optical fibers}
%
%
% author names and IEEE memberships
% note positions of commas and nonbreaking spaces ( ~ ) LaTeX will not break
% a structure at a ~ so this keeps an author's name from being broken across
% two lines.
% use \thanks{} to gain access to the first footnote area
% a separate \thanks must be used for each paragraph as LaTeX2e's \thanks
% was not built to handle multiple paragraphs
%
\author{I.~Roudas,~\IEEEmembership{Member,~IEEE}, J.~Kwapisz, and D.~A.~Nolan,~\IEEEmembership{Fellow,~OSA}}

\thanks{I.~Roudas is with the Department of Electrical and Computer Engineering, Montana State University, Bozeman, MT 59717, USA email: ioannis.roudas@montana.edu}  
\thanks{J.~Kwapisz is with the Department of Mathematical Sciences, Montana State University, Bozeman, MT 59717, USA email: jarek@math.montana.edu}
\thanks{D.~A.~Nolan is with the Science and Technology Division, Corning Research and Development Corporation, Corning, NY 14831, USA email: nolanda@corning.com}

% note the % following the last \IEEEmembership and also \thanks - 
% these prevent an unwanted space from occurring between the last author name
% and the end of the author line. i.e., if you had this:
% 
% \author{....lastname \thanks{...} \thanks{...} }
%                     ^------------^------------^----Do not want these spaces!
%
% a space would be appended to the last name and could cause every name on that
% line to be shifted left slightly. This is one of those "LaTeX things". For
% instance, "\textbf{A} \textbf{B}" will typeset as "A B" not "AB". To get
% "AB" then you have to do: "\textbf{A}\textbf{B}"
% \thanks is no different in this regard, so shield the last } of each \thanks
% that ends a line with a % and do not let a space in before the next \thanks.
% Spaces after \IEEEmembership other than the last one are OK (and needed) as
% you are supposed to have spaces between the names. For what it is worth,
% this is a minor point as most people would not even notice if the said evil
% space somehow managed to creep in.

% The paper headers
\markboth{JOURNAL OF LIGHTWAVE TECHNOLOGY, \today} %February~2018}%
{Shell \MakeLowercase{\textit{I. Roudas et al.}}: Optimal launch states for the measurement of principal modes in optical fibers}
% The only time the second header will appear is for the odd numbered pages
% after the title page when using the twoside option.
% 
% *** Note that you probably will NOT want to include the author's ***
% *** name in the headers of peer review papers.                   ***
% You can use \ifCLASSOPTIONpeerreview for conditional compilation here if
% you desire.

% If you want to put a publisher's ID mark on the page you can do it like
% this:
%\IEEEpubid{0000--0000/00\$00.00~\copyright~2015 IEEE}
% Remember, if you use this you must call \IEEEpubidadjcol in the second
% column for its text to clear the IEEEpubid mark.

% use for special paper notices
%\IEEEspecialpapernotice{(Invited Paper)}

% make the title area
\maketitle

% As a general rule, do not put math, special symbols or citations
% in the abstract or keywords.
\begin{abstract}
Modal dispersion characterization of multimode optical fibers can be performed using the recently-proposed mode-dependent signal delay method. This method consists of sending optical pulses using different combinations of modes though the multimode optical fiber and measuring the mode group delay at the fiber output. From these measurements, it is possible to estimate the modal dispersion vector, the principal modes, and their corresponding differential mode group delays. 

In this paper, we revise and extend the theoretical framework of the mode-dependent signal delay method to include the impact of receiver noise and mode-dependent loss. We compute optimal launch modes, minimizing the noise error in the estimation of the fiber modal dispersion vector. We show that, for a 40-mode fiber, the electronic signal-to-noise ratio (SNR) is improved asymptotically by almost 6 dB compared to conventional mode combinations. 
\end{abstract}

% Note that keywords are not normally used for peerreview papers.
\begin{IEEEkeywords}
Modal dispersion, multimode fiber characterization.
\end{IEEEkeywords}

% For peer review papers, you can put extra information on the cover
% page as needed:
% \ifCLASSOPTIONpeerreview
% \begin{center} \bfseries EDICS Category: 3-BBND \end{center}
% \fi
%
% For peerreview papers, this IEEEtran command inserts a page break and
% creates the second title. It will be ignored for other modes.
\IEEEpeerreviewmaketitle

\section{Introduction}
% The very first letter is a 2 line initial drop letter followed
% by the rest of the first word in caps.
% 
% form to use if the first word consists of a single letter:
% \IEEEPARstart{A}{demo} file is ....
% 
% form to use if you need the single drop letter followed by
% normal text (unknown if ever used by the IEEE):
% \IEEEPARstart{A}{}demo file is ....
% 
% Some journals put the first two words in caps:
% \IEEEPARstart{T}{his demo} file is ....
% 
% Here we have the typical use of a "T" for an initial drop letter
% and "HIS" in caps to complete the first word.
\IEEEPARstart{I}{nternet} traffic is expected to grow steadily in the near future \cite{Winzer:Parallelism}. For instance, Cisco predicts a 24\% compound annual growth of   global data traffic through the Internet from 2016 to 2021 \cite{Cisco}. If this trend persists over longer periods of time, it could eventually lead to a capacity shortage in the global fiber-optic network \cite{Chraplyvy}.

To address this challenge, researchers have been considering for some time the introduction of new fiber technologies that can support petascale data traffic per link in a cost-efficient way. For instance, it is possible to increase link capacity by using spatial division multiplexing (SDM), i.e., parallel transmission of optical data streams over disjointed spatial paths \cite{Winzer} provided by multimode and multicore optical fibers (jointly abbreviated below by the composite acronym SDM MMFs) \cite{Mizuno}. So far, it has been shown, both theoretically and experimentally, that strongly-coupled, single-mode, homogeneous multicore fibers exhibit a slight performance advantage over single-mode fiber (SMF) bundles \cite{Mumtaz}-\cite{Ryf:ECOC17}. Whether this argument alone is sufficient for the adoption of such optical fibers by the telecommunications market remains to be seen. 

It is assumed here that SDM MMFs will be eventually adopted in long-haul optical communications systems. A distinct feature of SDM MMFs for long-haul applications is that their modal dispersion (MD) should be very low, ideally comparable to the levels of polarization-mode dispersion (PMD) of SMFs, in order to facilitate digital signal processing at the coherent optical receiver \cite{Winzer}.  

By analogy with PMD \cite{Poole}-\cite{Kogelnik}, MD in long SDM MMFs can be  described by a set of propagation modes called principal modes (PMs) and by their corresponding differential mode group delays (DMGDs) compared to the average mode group delay \cite{Ho}. These quantities can be geometrically represented by a vector in a generalized Stokes space called MD vector \cite{Antonelli}, \cite{Roudas:PJ17}, which is a direct extension of the PMD vector in the conventional Stokes space \cite{Poole}-\cite{Kogelnik}. 

Taking further advantage of the similarity between MD and PMD, it is possible to modify previously-proposed PMD measurement techniques \cite{Kogelnik} and use them for MD characterization \cite{Ryf:BookChapter}-\cite{Carpenter}.  For instance, the polarization-dependent signal delay method can be used for the measurement of the PMD vector of SMFs \cite{Nelson}. The recently-proposed mode-dependent signal delay method \cite{Milione} is a generalization of the polarization-dependent signal delay method that can be used for the measurement of the MD vector of SDM MMFs. Namely, it relies on the determination of the components of the MD vector by launching optical pulses corresponding to different combinations of modes at the fiber input and measuring the corresponding group delays at the fiber output \cite{Milione}. From the MD vector, one can construct a Hermitian matrix, called the group-delay operator, and determine the PMs and the DMGDs from its eigenvectors and eigenvalues, respectively \cite{Antonelli}, \cite{Roudas:PJ17}.

An important question that is left unanswered in  previous articles on the mode-dependent signal delay method \cite{Milione}, \cite{Yang} is which  launch modes  must be used to measure the MD vector. Let $N$ be the number of spatial and polarization modes in the SDM  MMF under test. Then, the dimensionality of the generalized Stokes space is $N^2-1$ \cite{Antonelli}, \cite{Roudas:PJ17}. We need to choose ${N^2-1}$ different combinations of launch modes to determine the ${N^2-1}$ components of the MD vector. A set of ${N^2-1}$ launch states  corresponding to ${N^2-1}$ orthonormal vectors in Stokes space would constitute the best coordinate system for conducting such measurements. For $N>2$, however, it is impossible to find $N^2-1$ launch mode combinations corresponding to $N^2-1$ orthonormal vectors in the generalized Stokes space due to the incomplete coverage of the Poincar\'{e} sphere with valid states \cite{Antonelli}, \cite{Roudas:PJ17}. 

Milione et al. do not address this issue \cite{Milione}, while Yang and Nolan  \cite{Yang} propose a  set of modes that is a generalization into higher dimensions of the linear horizontal, linear 45 deg, and right-circular states of polarization used for  measurements  conducted in the conventional three-dimensional Stokes space. A weakness of the mode set proposed by \cite{Yang} is that, for $N>2$, it leads to suboptimal performance in the presence of receiver noise, as shown  in Sec. \ref{sec:results}. 

In the present paper, we analyze the impact of noise on the MD vector characterization process  performed in the mode-dependent signal delay method. Our analysis reveals that using a set of launch modes corresponding to an oblique vector set in the generalized Stokes space always magnifies the error in the estimation of the MD vector. Therefore, we should seek launch states that correspond to maximally-orthogonal Stokes vectors in order to minimize this error. In the sections that follow, we propose two numerical optimization algorithms based on the gradient descent method \cite{Boyd} that search the generalized Stokes space for $N^2-1$ maximally-orthogonal vectors corresponding to feasible launch states.

Furthermore, the presence of mode-dependent loss (MDL) is not considered in  the original articles on the mode-dependent signal delay method \cite{Milione}, \cite{Yang}. This is a significant omission in the sense that long SDM MMFs always exhibit a certain amount of MDL. In the present paper, we show that the mode-dependent signal delay method can be modified to characterize both MD and MDL simultaneously. 

In the following, we derive, from first principles, the fundamental equations for the determination of the MD and MDL vectors by using the mode-dependent signal delay method in the presence of additive white Gaussian noise, MD, and MDL  (Sec. \ref{sec:Fundamental_concepts}--Sec. \ref{sec:MDandMDL}). Practical considerations that might affect the accuracy  of the mode-dependent signal delay method  are discussed in Sec. \ref{Practical_considerations}. The remainder of this paper is devoted to the description of two optimization algorithms based on the gradient descent method for the selection of maximally-orthogonal launch states (Sec. \ref{sec:optimization1} and Sec. \ref{sec:algorithms}, respectively). Using these algorithms, we compute optimal  sets of  launch modes for up to $N=40$ that maximize the signal-to-noise ratio (SNR) at the direct-detection receiver and enhance the accuracy of the mode-dependent signal delay method (Sec. \ref{sec:results}). For example, for a 40-mode SDM MMF, we show that the optimal mode combinations improve the noise performance of the mode-dependent signal delay method by almost 6 dB compared to the set of modes proposed by Yang and Nolan \cite{Yang}. We also compare the noise performance of the proposed optimal mode combinations to vector sets often used for measurements in quantum mechanics, i.e., symmetric, informationally complete, positive operator valued measure (SIC-POVM) vectors  \cite{Fuchs} and vectors selected from mutually unbiased bases (MUBs) \cite{Bandyopadhyay} (see Appendix \ref{sec: Appendix}). We show that the proposed optimal mode combinations exhibit superior noise performance asymptotically by 3 dB compared to the aforementioned vector sets.

\section{Mathematical model \label{sec:model}}

\subsection{Fundamental concepts \label{sec:Fundamental_concepts}}

An intuitive way to quantify the modal dispersion of an optical fiber is through the time-of-flight of an optical pulse traveling along the fiber. The optical pulse experiences different delays depending on the combination of modes that are excited at the fiber input, the group velocity differences among the propagation modes, and the mode coupling at various points inside the fiber. The received pulse is a mixture of a multitude of components that arrive at the receiver at slightly different times and interfere constructively or destructively. The extraction of information regarding the fiber modal dispersion  from pulse delays, exclusively, constitutes the cornerstone of the mode-dependent signal delay method. 

In the absence of MDL (see Sec. \ref{sec:BasicMDequations}), an arbitrary launch mode can always be written as a linear combination of principal modes  at the fiber input. Furthermore, the group delay of a narrowband optical pulse at the fiber output can be written as a weighted superposition of the group delays experienced by the principal modes. The weights of the superposition are  functions of the excitations of the input principal modes. Formally, this weighted superposition can be written in a concise form in Stokes space as the dot product between the input MD vector representing the MD of the optical fiber and a unit Stokes vector representing the launch combination of modes (see expression (\ref{eq:MGD})).

The basic idea of the mode-dependent signal delay method is to identify the components of the MD vector. This can be accomplished by launching optical pulses at the fiber input corresponding to different combinations of modes  and measuring the corresponding group delays at the fiber output \cite{Milione}. Assume that different pulses excite mode combinations corresponding to vectors that linearly span Stokes space. Then, the group delays experienced by different pulses are enough to recover the input MD vector. 

If the above vectors are linearly independent, this corresponds to a decomposition of the MD vector  into the basis of the Stokes vectors representing the launch combinations of modes. We can  write the components of the Stokes vectors in the form of a matrix, the coefficient matrix of the linear system (see expression (\ref{eq:coefficientMatrix})). We can also write the group delays corresponding to different mode combinations as a column vector (see expression (\ref{eq:dmgds})). The components of the MD vector can then be recovered by solving a set of linear equations (see expression (\ref{eq:MDvectorCalculation})). 

If the Stokes vectors representing the mode combinations are pairwise orthogonal, the solution of the aforementioned set of linear equations is less affected by the unavoidable presence of random perturbations in the measurements (e.g., thermal noise, errors in the settings of the mode converter). Otherwise, matrix inversion leads to error amplification.

We follow a similar procedure to the one described above in order to measure the MDL of an optical fiber (see Sec. \ref{sec:MDandMDL}). For MDL characterization, we perform measurements  of the average output power of CW optical waves. A continuous optical wave can always be decomposed into a superposition of principal attenuation modes  at the fiber input. The attenuation of the optical wave can be written as a weighted superposition of the attenuations experienced by the principal attenuation modes. The weights of the superposition are  functions of the excitations of the input principal attenuation modes. Formally, this weighted superposition can be written in a concise form in Stokes space as the dot product between the input MDL vector, representing the MDL of the optical fiber, and a unit Stokes vector representing the launch combination of modes (see expression (\ref{eq: averagePowerFinal})).

In summary, the mode-dependent signal delay method can be used to measure both MD and MDL. In the joint presence of MD and MDL, we use a more sophisticated, multi-step version of the mode-dependent signal delay method, involving two separate sets of measurements, for MDL and MD, respectively, and an intermediate MDL equalization phase (see Sec. \ref{sec:MDandMDL}). First, we carry out MDL characterization by sending CW light into the fiber under test and measuring the average output power, as described above. Measurements for different launch mode combinations enable us to retrieve the individual components of the MDL vector. Once the MDL vector is known, we can optically compensate for the optical fiber's MDL at the fiber input by adjusting the settings of the mode converter. Finally, the  MD vector of the MDL-equalized optical fiber can be determined by launching optical pulses at the fiber input corresponding to different combinations of modes  and measuring the corresponding group delays at the fiber output, as described in the beginning of this subsection. In both sets of measurements, for MDL and MD characterization, we can use the same set of launch modes corresponding to maximally-orthogonal Stokes vectors in order to minimize noise errors.

On a final note, it is worth explaining the motivation behind using the generalized Stokes formalism for the analysis of the experimental measurements provided by the mode-dependent signal delay method. The familiar three-dimensional (3D) Stokes space is traditionally used to provide a geometrical representation of states of polarization (SOPs) in terms of  real vectors. This is useful for visualization of the transformations of SOPs by optical systems. For instance, the spatial evolution of the SOP of a monochromatic optical plane wave traveling through a birefringent SMF can be graphically depicted by a trajectory on the surface of the Poincar\'{e} sphere.
The Stokes space formalism may be extended to higher dimensions \cite{Antonelli}, \cite{Roudas:PJ17} to describe geometrically mode combinations and their evolution during propagation through "modally-birefringent" SDM MMFs, but then the intuitive visual appeal of the conventional 3D Stokes representation is lost. Nevertheless, from  a modeling perspective, there is still an incentive for using the generalized Stokes space instead of the generalized Jones space. The reason is that measurable quantities provided by direct-detection receivers can be conveniently expressed as dot products of generalized Stokes vectors. Therefore, the use of the generalized Stokes space enables us to perform analytical calculations solely in terms of real vectors, avoiding to employ complex matrices in the generalized  Jones space.

\subsection{Experimental process}
\begin{figure}[!ht]
\centering
	\centering
\begin{subfigure}[b]{0.4\textwidth}
	\includegraphics[width=1\linewidth]{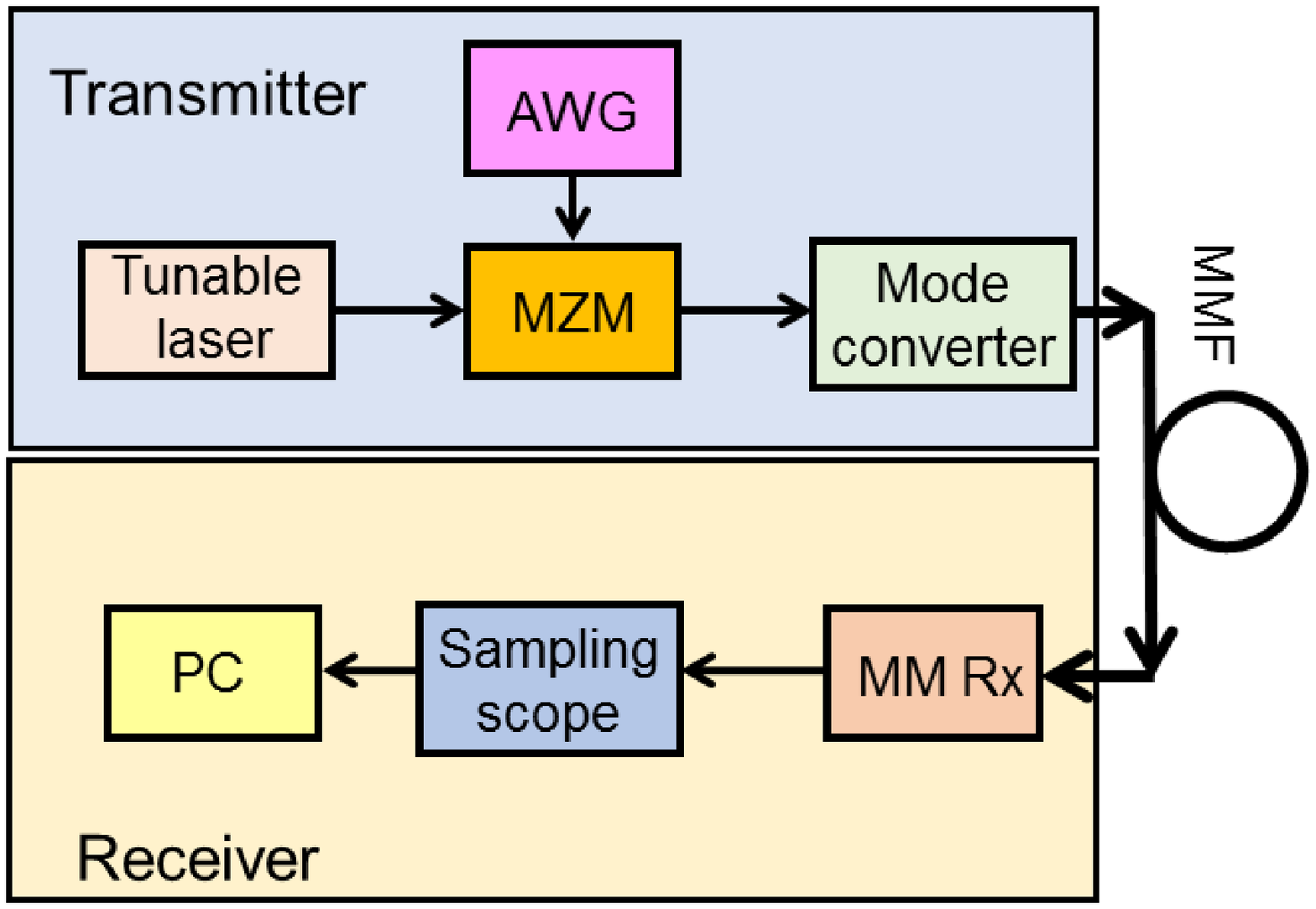}%{ImprovedSetup}
	\caption{}
	\label{fig:expSetup}
\end{subfigure}

\begin{subfigure}[b]{0.3\textwidth}
	\includegraphics[width=1\linewidth]{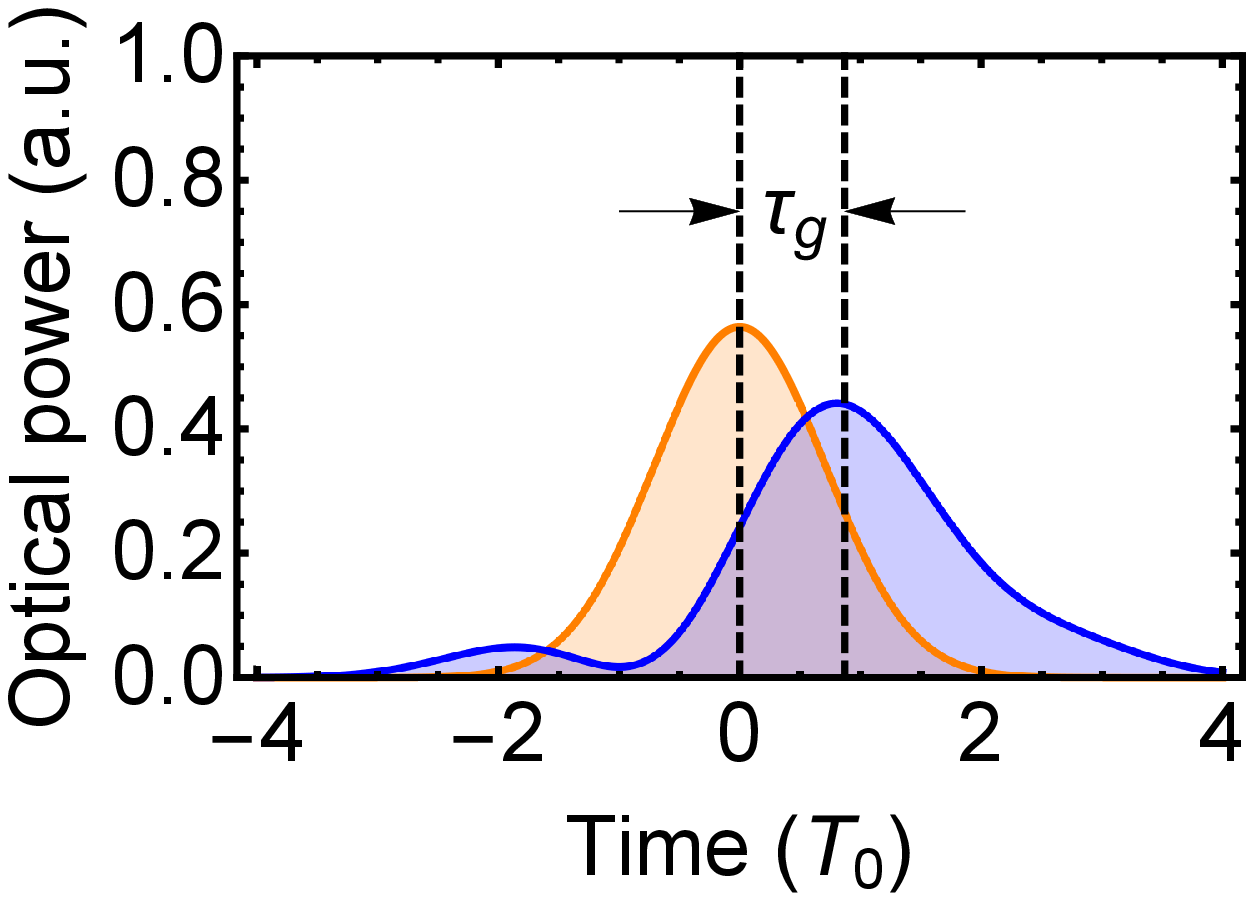}
	\caption{}
	\label{fig:CartoonPulses}
\end{subfigure}

\caption{(a) Experimental setup for MD characterization of SDM MMFs using the mode-dependent signal delay method (Abbreviations: AWG: arbitrary waveform generator, MZM: Mach-Zehnder modulator, MMF: multimode fiber, MM Rx: multimode receiver, PC: computer); (b) Input Gaussian pulse with unit energy (orange line) and output pulse (blue line). The origin of the time axis coincides with the center of the input pulse. (Symbols: $T_{0}$=half-width at 1/e power point \cite{Agrawal}; $\tau_{g}$=group delay of the output pulse).}
%\label{fig:preliminary}
\end{figure}

\noindent The experimental setup used in the mode-dependent signal delay method is shown in Fig. \ref{fig:expSetup}. The transmitter consists of a tunable laser externally modulated by a Mach-Zehnder modulator (MZM) using an arbitrary waveform generator (AWG). ~This configuration provides narrowband, Gaussian, transform-limited, optical pulses with adjustable duration and repetition rates. The laser scans the whole frequency band of interest in steps larger than the pulse bandwidth. The purpose of the mode converter is to generate arbitrary spatial and polarization mode combinations. Applying appropriate mode excitations, one can launch optical pulses which experience different propagation delays. After the SDM MMF under test, the optical pulses are detected, sampled using a real-time oscilloscope, and stored in a PC. Off-line processing can be used for the evaluation of the PMs and the DMGDs at each frequency.

Indicative drawings of the input and output pulses are shown in Fig.~\ref{fig:CartoonPulses} (orange and blue lines, respectively). In the following subsections, we link the group delay ${\tau }_g\ $to the MD vector and the unit Stokes vector representing the launch combination of modes.

\noindent 

\subsection{Basic notation}
\noindent Throughout this article, we use the notation conventions introduced by \cite{Gordon}: Dirac's bra-kets refer to unit Jones vectors. Unit Stokes vectors are denoted by carets, while non-unit Stokes vectors are denoted by arrows. Matrices are designated by uppercase boldface letters.

We can write any $N\times N$ matrix ${\bf M}$ as a linear superposition of the identity matrix ${\bf I}$ and the $N^2-1$ generalized Gell-Mann matrices  ${\boldsymbol{\mathrm{\Lambda }}}_i, i=1,\ldots,N^2-1 $ \cite{Roudas:PJ17}. We denote by $\boldsymbol{\mathrm{\Lambda }}$ the Gell-Mann column vector 
\begin{equation}
\boldsymbol{\mathrm{\Lambda }}\df {\left[{\boldsymbol{\mathrm{\Lambda }}}_1,\dots ,{\boldsymbol{\mathrm{\Lambda }}}_{N^2-1}\right]}^T,  
\end{equation}
where the superscript $T$ denotes the transpose of a matrix.

The unit Stokes vector $\hat{s}\ $corresponding to the unit Jones vector $\left.|s\right\rangle $ is defined as \cite{Roudas:PJ17} 
\begin{equation}
\hat{s}\df C_N \left\langle s\mathrel{\left|\vphantom{s \boldsymbol{\mathrm{\Lambda }} s}\right.\kern-\nulldelimiterspace}\boldsymbol{\mathrm{\Lambda }}\mathrel{\left|\vphantom{s \boldsymbol{\mathrm{\Lambda }} s}\right.\kern-\nulldelimiterspace}s\right\rangle ,
\label{eq:StokesVectors}
\end{equation}
where $C_N$ denotes the normalization coefficient \cite{Roudas:PJ17}
\begin{equation}
C_N\df \sqrt{N/\left[2\left(N-1\right)\right]}.
\end{equation}

For each Jones vector $\left.|s\right\rangle $, we can define the associated projection operator $| s \rangle \langle s |$, which represents a mode filter, i.e., the equivalent of a polarizer in the two-dimensional case. The projection operator can be expressed in terms of the identity matrix and the generalized Gell-Mann matrices \cite{Roudas:PJ17}
\begin{equation}
\label{jonesToop:eq}
| s \rangle \langle s | = \frac{1}{N}{\bf I }+\frac{1}{2 C_N} \hat{s}\cdot \bLambda.
\end{equation}

In (\ref{jonesToop:eq}), we used the dot product of the Stokes vector $\hat{s}$ with the Gell-Mann vector $\bLambda$. This is defined as the sum of the products of the corresponding entries 
\begin{equation}
\hat{s}\cdot \bLambda:=\sum_{i=1}^{N^2-1}s_i \bLambda_i
\end{equation}

Finally, we will use, without proof, a relationship between the inner product
% $\sh_j \cdot \sh_k := \sh_j\cdot \sh_k = \sum_v \hat{s}_{jv} \cdot \hat{s}_{kv}$
in Stokes space and the inner product in Jones space derived in \cite{Roudas:PJ17}:
\beq
\label{JonesOpTostokesInnerProd:eq}
% \hat{s}_j \cdot \hat{s}_k  %= \left( [s_j] \middle| [s_k] \right)
\sh_j\cdot \sh_k 
= 2C_N^2 \left[ | \langle s_j |  s_k \rangle |^2 - \frac{1}{N}  \right].
\eeq

\subsection{MD vector definition}

\noindent First, we consider the ideal case of an \textit{N}-mode SDM MMF with negligible MDL. The fiber transfer function can be described by a generalized Jones unitary matrix $\boldsymbol{\mathrm{U}}\left(\omega \right).$  We define the  input group-delay operator $i{\boldsymbol{\mathrm{U}}}^{\dagger }\left(\omega \right){\boldsymbol{\mathrm{U}}}_{\omega }\left(\omega \right)$, where the subscript $\omega $ denotes differentiation with respect to the angular frequency and   a raised dagger denotes the adjoint matrix.    
 
The input group-delay  operator can be represented in the basis of the identity matrix and the generalized Gell-Mann matrices  
\begin{equation}
\begin{array}{c}
i{\boldsymbol{\mathrm{U}}}^{\dagger }\left(\omega \right){\boldsymbol{\mathrm{U}}}_{\omega }\left(\omega \right)\df {\tau }_0 \left(\omega \right){\bf I}+\frac{1}{2 C_N}{\vec{\tau }}_s\left(\omega \right)\cdot\boldsymbol{\mathrm{\Lambda }},\end{array}
\label{eq:MDvectorDefinition}
\end{equation}
where  ${\tau }_0 \left(\omega \right)$ is the average group delay and ${\vec{\tau }}_s$ is the input MD vector \cite{Roudas:PJ17}.   

The eigenvalues and eigenvectors of the operator ${1}/{(2 C_N)}{\vec{\tau }}_s\left(\omega \right)\cdot\boldsymbol{\mathrm{\Lambda }}$ are the DMGDs ${\ \tau }_i\left(\omega \right)$ and the input PMs $\left.|p_i\left(\omega \right)\right\rangle $,  ${i}=1,{\dots}, N$. We can write the eigenvalue equation
\begin{equation}
\begin{array}{c}
\frac{1}{2 C_N}{\vec{\tau }}_s\left(\omega \right)\cdot\boldsymbol{\mathrm{\Lambda }} \left|\left.p_i\left(\omega \right)\right\rangle ={\tau }_i\left(\omega \right)\right|\left.p_i\left(\omega \right)\right\rangle. \end{array}
\label{eq:GDoperator}
\end{equation}

\subsection{Information provided by the input MD vector}

\noindent Before we delve further into the mode-dependent signal delay method, it is worth investigating whether the MD vector ${\vec{\tau }}_s \left(\omega \right)$ and the fiber transfer matrix ${\boldsymbol{\mathrm{U}}}\left(\omega \right)$ are really equivalent representations of the optical fiber modal dispersion. 

From the definition of the input MD vector (\ref{eq:MDvectorDefinition}), setting the average group delay ${\tau }_0=0 $ for simplicity, we obtain
\begin{equation} \label{1)} 
\begin{array}{c}
i{\boldsymbol{\mathrm{U}}}^{\dagger }\left(\omega \right){\boldsymbol{\mathrm{U}}}_{\omega }\left(\omega \right)= \frac{1}{2 C_N}{\vec{\tau }}_s\left(\omega \right)\cdot\boldsymbol{\mathrm{\Lambda }}\end{array}. 
\end{equation} 

Acting on both sides of this expression from the left with $-i {\bf U}(\omega )$ yields a $N\times N$ homogeneous system of coupled first-order ordinary differential equations
\begin{equation} \label{2)} 
{\bf U}_{\omega } (\omega )={\bf U}(\omega )\left[-\frac{i}{2C_{N} } \vec{\tau }_{s}^{} \left(\omega \right)\cdot \boldsymbol{\mathrm{\Lambda }} \right]. 
\end{equation} 

Assume that $\vec{\tau }_{s}^{} \left(\omega \right)$ is constant in the interval $\left[\omega_0, \omega_0+\delta\omega\right]$. Then, the solution of the above system in matrix form is
\begin{equation} \label{3)} 
{\bf U}\left(\omega_0+\delta\omega\right)={\bf U}\left(\omega_0\right)\exp\left[{-\frac{i}{2C_{N} } \vec{\tau }_{s}^{} \left(\omega_0\right)\cdot \boldsymbol{\mathrm{\Lambda }} \delta\omega}\right] . 
\end{equation} 

Notice that knowledge of $\vec{\tau }_{s}^{} \left(\omega_0 \right)$ is insufficient to fully determine the fiber transfer matrix ${\bf U}(\omega_0+\delta\omega).$ This would require knowledge of ${\bf U}(\omega_0)$ as well, which is not provided by the mode-dependent signal delay method.

Measurement of the input MD vector $\vec{\tau }_{s}^{} \left(\omega \right)$  enables the determination of the DMGDs and the input PMs, exclusively, as shown in (\ref{eq:GDoperator}). In this respect,  the input MD vector $\vec{\tau }_{s}^{} \left(\omega \right)$ indeed encapsulates the modal dispersion of the fiber. In contrast, the fiber transfer matrix ${\bf U}(\omega )$ inherently contains additional information, e.g., one can also determine the output PMs, as well.

Based on the above discussion, we conclude that the mode-dependent signal delay method does not provide the same information as alternative methods measuring the fiber transfer matrix, e.g., swept wavelength interferometry \cite{Ryf:BookChapter}-\cite{Carpenter}.

\subsection{Linking the input MD vector to the pulse group delay \label{sec:BasicMDequations}}
\noindent Following the methodology of \cite{Gordon} and \cite{Milione}, assuming perfectly coherent signals, we can  prove that the group delay ${\tau }_g$ of an optical pulse with carrier angular frequency $\omega$ propagating through the optical fiber under test is related to the  input MD vector ${\vec{\tau }}_s$ and the Stokes vector $\hat{s}$ representing the combination of launch modes \cite{Roudas:PJ17}
\begin{equation}
\begin{array}{c}
{\tau }_g={\tau }_0+ \frac{1}{2 C_N^2}\left\langle {\vec{\tau }}_s \right\rangle \cdot\hat{s}, \end{array}
\label{eq:MGD}
\end{equation}
where all quantities depend on the carrier angular frequency $\omega$ but we omitted this dependence for notational simplicity.

In (\ref{eq:MGD}), the group delay ${\tau }_g$ is defined as the first moment in time \cite{Gordon}
\begin{equation}
{\tau _g} \df \frac{1}{\bar{E}}\int\limits_{ - \infty }^\infty  {t{{\bf{E}}_0}{{\left( t \right)}^\dag }{{\bf{E}}_0}\left( t \right)dt},
\label{eq:tg}
\end{equation}
where ${\bf{E}}_0(t)$ is the electric field at the fiber output and $\bar{E}$ is the average energy of the received pulse  
\beq
\bar{E} \df \int\limits_{ - \infty }^\infty  {{{\bf{E}}_0}{{\left( t \right)}^\dag }{{\bf{E}}_0}\left( t \right)dt} .
\eeq

Finally, in (\ref{eq:MGD}), we defined the spectrally-averaged input MD vector $\left\langle {\vec{\tau }}_s\right\rangle $ as
%\beq
%{\left\langle {{{\vec \tau }_s}\left( \omega  \right)} \right\rangle  \df \frac{\int\limits_{ - \infty }^\infty  {{{\vec \tau }_s}\left( \omega  \right){{\left| {G\left( \omega  \right)} \right|}^2}d\omega } }{{\int\limits_{ - \infty }^\infty  {{{\left| {G\left( \omega  \right)} \right|}^2}d\omega }  }}},
%\eeq
\beq
{\left\langle {{{\vec \tau }_s}} \right\rangle  \df \frac{1}{{2\pi \bar{E}}}\int\limits_{ - \infty }^\infty  {{{\vec \tau }_s}\left( \omega'  \right){{\left| {G\left( \omega'  \right)} \right|}^2}d\omega' } },
\eeq
where $G\left( \omega  \right)$ is the input pulse spectrum.

Our expression (\ref{eq:MGD}) differs from the initial expression (16) of Milione et al. \cite{Milione} on two important points: the input MD vector ${\vec{\tau }}_s\ $is spectrally-averaged and there is a corrective multiplicative factor of ${1}/{2 C_N^2}$ in front of the inner product $\left\langle {\vec{\tau }}_s \right\rangle \cdot\hat{s}$. 

In the following, we assume that we use optical pulses with sufficiently narrow spectrum around the carrier angular frequency $\omega$ so that $\left\langle {\vec{\tau }}_s\right\rangle \simeq {\vec{\tau }}_s.\ $

\subsection{MD vector estimation\label{sec:MDvectorEstimation}}

First, the average group delay ${\tau }_0$ in (\ref{eq:MGD}) can be estimated by using the following procedure: We launch pulses corresponding to \textit{N} arbitrary orthogonal states in Jones space $\ket{s_{0,i}}$ and measure their group delays ${\tau }_{0g,i}$, $i=1,\ldots,N$. We know that $N$ orthonormal vectors in Jones space are mapped into Stokes vectors $\hat{s}_{0,i}$ that form the vertices of a $(N-1)$-dimensional regular simplex \cite{Aerts}. This implies that $\hat{s}_{0,i}$ sum to zero \cite{Roudas:PJ17}
\beq\sum_{i = 1}^N {{{\hat s}_{0,i}}}=0.\eeq

Taking the average of the corresponding group delays ${\tau }_{0g,i}$   \cite{Roudas:PJ17}, expression (\ref{eq:MGD}) yields
\begin{equation}
{\tau }_0=\frac{1}{N}\sum_{i = 1}^N {\tau _{0g,i}^{}}.
\label{eq: DMDzero}
\end{equation}

Subsequently, we launch ${(N}^2-1)$ linearly independent input states in Stokes space and measure the corresponding group delays ${\tau }_{g_i}$, $i=1,{\dots},N^2-1$. Expression (\ref{eq:MGD}) can be used to form a ${(N}^2-1)$$\times$${(N}^2-1)$  system of linear equations. We can represent these equations in matrix form. First, we define the coefficient matrix
\begin{equation}
{\boldsymbol{\mathrm{S}}}  \df  \left[{\hat{s}}_1,\dots {,\hat{s}}_{N^2-1}\right]^{{T}}.
\label{eq:coefficientMatrix}
\end{equation}
Notice that the columns of ${\bf S}^T$ are the launch states represented by the Stokes vectors ${\hat{s}}_i$, ${i}=1,{\dots}, N^2-1$.

Then, we define the column vector of the DMGD's 
\begin{equation}
{\boldsymbol{\mathrm{T}}}_{\mathrm{g}}  \df  2 C_N^2 {\left[\mathrm{\ }{\tau }_{g,1}-{\tau }_0,\dots ,\mathrm{\ }{\tau }_{g,N^2-1}-{\tau }_0\right]}^T.
\label{eq:dmgds}
\end{equation} 

Finally, the matrix representation of the system of equations is written as
\begin{equation}
{\boldsymbol{\mathrm{S}}}{\vec{\tau }}_s= {\boldsymbol{\mathrm{T}}}_{\mathrm{g}}.
\end{equation}

Thus, the MD vector ${\vec{\tau }}_s$ is given by  
\begin{equation}
	{\vec{\tau }}_s= {\boldsymbol{\mathrm{S}}}^{\boldsymbol{-}\boldsymbol{1}}{\boldsymbol{\mathrm{T}}}_{\mathrm{g}}.
	\label{eq:MDvectorCalculation}
	\end{equation}

For ${\bf S}$ to be invertible, its determinant must be nonzero, $\text{det}({\bf S})\neq 0$, so ${\hat{s}}_i$, ${i}=1,{\dots}, N^2-1$, have to be linearly independent.

\subsection{Noise modeling}
\noindent Let us investigate the effect of the thermal noise $n(t)$ of the direct-detection receiver on the measurement of the DMGD's.

The error in the estimate of the group delay ${\tau }_g$ is
\begin{equation}
{\delta \tau_{g} =\frac{1}{R_d\bar{E}}\int _{-T/2}^{T/2}tn(t)dt } ,
\label{eq:MGDerror}
\end{equation}
where $R_d$ denotes the responsivity of the photodiode and $T$ denotes the integration time for the computation of the group delay ${\tau }_g$.

We assume that the thermal noise at the direct-detection receiver can be modeled as an additive white Gaussian noise (AWGN) with zero mean and  autocorrelation function given by \cite{Proakis}
\begin{equation}
R(t,t'):={E\left\{n(t)n(t')\right\}=\frac{N_{0} }{2} \delta \left(t-t'\right)},
\label{eq:autocorrelationThermalNoise}
\end{equation}
where the operator $E\left\{.\right\}$ denotes the expected value, ${N_{0} }/{2}$ denotes the power spectral density of the noise, and $\delta(t)$ is the Dirac delta function.

The mean of $\delta \tau_{g}$ is calculated by taking the expectation of both sides of  (\ref{eq:MGDerror}) \cite{Proakis}
\begin{equation}
\mu _{\delta \tau_{g} } :={E\left\{\delta \tau_{g} \right\}=\frac{1}{R_d\bar{E}}\int _{-T/2}^{T/2}tE\left\{n(t)\right\}dt =0}.
\label{eq:mean}
\end{equation}

The variance of $\delta \tau_{g}$ is given by \cite{Proakis}
\begin{equation}
{\sigma _{\delta \tau_{g} }^{2} :=\frac{1}{R_d^2\bar{E}^2}\int _{-T/2}^{T/2}\int _{-T/2}^{T/2}tt'E\left\{n(t)n(t')\right\}dtdt'  }.
\label{eq:varianceDeltaTg}
\end{equation}

Upon substituting (\ref{eq:autocorrelationThermalNoise}) into (\ref{eq:varianceDeltaTg}), we find
\begin{equation}
{\sigma _{\delta \tau_{g} }^{2}=\frac{N_{0} }{2R_d^2\bar{E}^2} \int _{-T/2}^{T/2}t^{2} dt =\frac{N_{0} T^{3} }{24R_d^2\bar{E}^2} }.
\label{eq:NoiseVarianceFinalExpression}
\end{equation}

The presence of thermal noise at the individual measurements ${\tau }_{g,i}$, ${i}=1,{\dots}, N^2-1$  can lead to a random offset ${\delta\boldsymbol{\mathrm{T}}}_{\mathrm{g}}\boldsymbol{\ }$ in the estimation of the DMGD matrix ${\boldsymbol{\mathrm{T}}}_{\mathrm{g}}$ in (\ref{eq:dmgds}). Consequently, there is an error in the estimate of ${\vec{\tau }}_s\left(\omega \right)$ in (\ref{eq:MDvectorCalculation}), namely 
\beq
\delta {\vec{\tau }}_s={\boldsymbol{\mathrm{S}}}^{-1}{\delta }{\boldsymbol{\mathrm{T}}}_{\mathrm{g}}={\boldsymbol{\mathrm{A}}}{\delta }{\boldsymbol{\mathrm{T}}}_{\mathrm{g}},
\label{eq:MDvectorError}
\eeq
where we set ${\bf A}\df{\bf S}^{-1}$ for brevity.

Taking the expectation of both sides of (\ref{eq:MDvectorError}) and substituting (\ref{eq:mean}) in (\ref{eq:MDvectorError}), we obtain the mean of $\delta {\vec{\tau }}_s$
\beq
\mu_{\delta {\vec{\tau }}_s}:={\rm E}\left\{ \delta {\vec{\tau }}_s \right\}={\boldsymbol{\mathrm{A}}}{\rm E}\left\{{\delta }{\boldsymbol{\mathrm{T}}}_{\mathrm{g}}\right\}={\bf 0}.
\label{eq:meandt}
\eeq

The covariance matrix of $\delta{\vec{\tau }}_s$ is given by
\begin{equation}
{\bf C}_{\delta {\vec{\tau }}_s}:={\rm E}\left\{\delta{\vec{\tau }}_s \delta{\vec{\tau }}_s^T \right\}={\bf A}{\bf C}_{\delta {\bf T}_{g}}{\bf A}^{T},
\label{eq:covariance}
\end{equation}
where we defined the covariance of $\delta {\bf T}_{g}$
\beq
{\bf C}_{\delta {\bf T}_{g}}\df{\rm E}\left\{\delta {\bf T}_{g}   { \delta {\bf T}_{g}^T } \right\}.
\eeq
 
We assume that the measurements of the components $ T_{g_i} $ of the column vector ${\boldsymbol{\mathrm{T}}}_{\mathrm{g}}\boldsymbol{\ }$ are performed sequentially. As a result, the noise realizations $n_i(t)$ in different measurements are independent and the random offsets   $\delta T_{g_i} $ are uncorrelated. Therefore, the covariance matrix of $\delta {\bf T}_{g}$ is diagonal with elements $\sigma _{\delta T_{g} }^{2}$
\beq
{\bf C}_{\delta {\bf T}_{g}}=\sigma _{\delta T_{g} }^{2}{\bf I},
\label{eq:covarianceTg}
\eeq
where
\begin{equation}
\sigma _{\delta T_{g} }^{2}  =  4 C_N^4 \sigma _{\delta \tau_{g} }^{2}.
\label{eq:varTg}
\end{equation} 

In (\ref{eq:varTg}), we assumed, for simplicity, that $\tau_{0}$ is a deterministic (i.e., a perfectly measured) quantity, as opposed to the group delays ${\tau }_{g_i}$, $i=1,{\dots},N^2-1,$ that are independent, identically-distributed random variables. The reason for this approximation is that the variance $\sigma _{\delta\tau_{0} }^{2}=\sigma _{\delta \tau_{g} }^{2}/N$, so it is much smaller than $\sigma _{\delta \tau_{g} }^{2}$ for large values of $N$. For small $N$'s, it might be necessary to repeat the process for measuring  $\tau_{0}$ a certain number of times  $k$ and average the results. Then, $\sigma _{\delta\tau_{0} }^{2}=\sigma _{\delta \tau_{g} }^{2}/(k N)$, so that $\sigma _{\delta\tau_{0} }^{2}$ is negligible compared to $\sigma _{\delta \tau_{g}}^2$ for large values of $k N$.

From (\ref{eq:covariance}), (\ref{eq:covarianceTg}), it follows immediately that
\begin{equation}
{\bf C}_{\delta {\vec{\tau }}_s}=\sigma _{\delta { T}_{g} }^{2} {\bf A}{\bf A}^{T}.
\label{eq:covariance2}
\end{equation}
 
The diagonal entries of the covariance matrix ${\bf C}_{\delta {\vec{\tau }}_s}$ are the variances of the components
of the random vector $\delta {\vec{\tau }}_s$. Thus, the variance of $\left\|\delta {\vec{\tau }}_s\right\|$ is given by
\begin{equation}
{\sigma }^2_{\left\|\delta {\vec{\tau }}_s\right\|}:=E\left\{\left\| \delta \vec{\tau }_{s}^{} \right\| ^{2} \right\}=E\left\{{\rm Tr}\left[{\bf C}_{\delta {\vec{\tau }}_s}\right]\right\},
\label{eq:varInterim}
\end{equation}
where the operator $\mytrace(.)$ denotes the trace of a matrix.

Interchanging the order of operators in (\ref{eq:varInterim}) and substituting  (\ref{eq:covariance2}),  we obtain
\begin{equation}
{\sigma }^2_{\left\|\delta {\vec{\tau }}_s\right\|}=\sigma _{\delta T_{g} }^{2} {\rm Tr}\left[{\bf A}{\bf A}^{T} \right],
\label{eq:MDVectorLengthVariance}
\end{equation}

We can calculate explicitly ${\rm Tr}\left[{\bf A}{\bf A}^{T} \right]$ from the singular value decomposition (SVD) of ${\bf S}$ \cite{Strang}
\begin{equation}
\mathbf {S} =\mathbf {W} {\boldsymbol {\Sigma }}\mathbf {V} ^{T},
\end{equation}
where ${\mathbf W}, {\mathbf V}$ are orthogonal matrices and ${\boldsymbol {\Sigma }}$ is a diagonal matrix with entries the singular values $\sigma_i$, ${i=1,\dots ,N^2-1},$ of $\mathbf {S}$.

Then, $\mathbf {A} =\mathbf {S}^{-1}=\mathbf {V} {\boldsymbol {\Sigma }}^{-1}\mathbf {W} ^{T}$ and we have
\begin{equation}
{\bf A}{\bf A}^{T} = {\mathbf V}{\boldsymbol {\Sigma }}^{-2}{\mathbf V}^T.
\end{equation}

Hence, we obtain a succinct expression for the trace
\begin{equation}
{\rm Tr}\left[{\bf A}{\bf A}^{T} \right]=\sum^{N^2-1}_{k=1}{{\sigma }^{-2}_{k }}.
\label{eq:TrAAt}
\end{equation}

Combining (\ref{eq:MDVectorLengthVariance}) and (\ref{eq:TrAAt}) yields 
\begin{equation}
{\sigma }^2_{\left\|\delta {\vec{\tau }}_s\right\|}={\sigma }^2_{{\delta T}_g}\sum_{k=1}^{N^2-1}{\sigma }_{k }^{-2}. 
\label{eq:MDvectorLengthVariance}
\end{equation}

To find a lower bound for (\ref{eq:MDvectorLengthVariance}), we use the fact that the arithmetic mean of the numbers $\sigma_k^{-2}$ is greater than or equal to the geometric mean of the set (arithmetic/geometric mean inequality)
\beq
\frac{1}{N^2-1} \sum_{k=1}^{N^2-1} \sigma_k^{-2} \geq \left( \prod_{k=1}^{N^2-1} \sigma_k^{-2} \right)^{1/(N^2-1)}.
\eeq
In addition,
\beq
\prod_{k=1}^{N^2-1} \sigma_k^{2} = {\rm det}({\SSS\SSS}^T) = {\rm det}({\SSS})^2 \leq 1,
\eeq
because  $|{\rm det}(\SSS)|$ is the volume of the $N^2-1$-dimensional parallelotope spanned by the unit Stokes vectors $\hat{s}_j$, which cannot exceed that of a cube.

Therefore,
\beq	
\sum_{k=1}^{N^2-1} \sigma_k^{-2}  \geq N^2-1.
\eeq

Notice that, if ${\bf S}$ is an orthogonal matrix, then so is ${\bf A}$ and ${\bf A}{\bf A}^{T} ={\bf I}$. Thus, we obtain the minimum estimated variance from (\ref{eq:MDVectorLengthVariance})
\begin{equation}
{\sigma }^2_{\left\|\delta {\vec{\tau }}_s\right\|}=(N^{2} -1)\sigma _{\delta T_{g} }^{2} .
\label{eq:minvar}
\end{equation}

In addition, since the off-diagonal elements of ${\bf A}{\bf A}^{T}$ are zero in this case, the components of $\delta \vec{\tau }_{s}^{} $ are uncorrelated and therefore independent. 

In contrast, if ${\bf S}$ is not an orthogonal matrix, then ${\sigma }^2_{\left\|\delta {\vec{\tau }}_s\right\|}>(N^{2} -1)\sigma _{\delta T_{g} }^{2},$ i.e., there is noise amplification compared to the previous case, which is due to the matrix inversion in (\ref{eq:MDvectorCalculation}). Moreover, the off-diagonal elements of ${\bf A}{\bf A}^{T}$ are nonzero and the components of $\delta \vec{\tau }_{s}^{} $ become correlated. 

\subsection{Joint measurement of MD and MDL \label{sec:MDandMDL}}

\noindent In our previous formulation, we assumed that MDL was negligible and that the fiber transfer matrix was unitary ${\bf U}(\omega ).$ This was an instructive special case.
In the presence of MDL, the fiber transfer matrix becomes non-unitary and is denoted by ${\bf H}(\omega )$. The output group delay operator then becomes $i{\bf H}_{\omega } \left(\omega \right){\bf H}^{-1} \left(\omega \right)$ \cite{Huttner}. The corresponding input group delay operator becomes $i{\bf H}^{-1} \left(\omega \right){\bf H}_{\omega } \left(\omega \right).$ The latter is a non-Hermitian matrix, so it has complex eigenvalues and non-orthogonal eigenvectors, in general \cite{Huttner}.

Following closely the methodology of \cite{Huttner}, we can represent the input group delay operator as a linear combination of the identity matrix and the Gell-Mann matrices
\begin{equation} \label{14)} 
i{\bf H}^{-1} \left(\omega \right){\bf H}_{\omega } \left(\omega \right)=\chi _{0}\left(\omega \right) {\bf I}+\frac{1}{2C_{N} } \vec{\chi }_{s} \left(\omega \right)\cdot \boldsymbol{\mathrm{\Lambda }} , 
\end{equation} 
where $\chi _{0}\left(\omega \right) ,\vec{\chi }_{s}\left(\omega \right)$  denote the complex mean group delay and the complex MD vector, respectively.

The input PMs are the eigenstates of the group-delay operator \cite{Huttner}
\begin{equation}
\begin{array}{c}
\frac{1}{2C_{N} } \vec{\chi }_{s} \left(\omega \right)\cdot \boldsymbol{\mathrm{\Lambda }}\left|\left.\chi_i\left(\omega \right)\right\rangle ={\chi }_i\left(\omega \right)\right|\left.\chi_i\left(\omega \right)\right\rangle, \end{array}
\label{eq:GDoperator2}
\end{equation} 

The DMGDs are given by \cite{Huttner}
\begin{equation} \label{7)} 
\tau _{i} \left(\omega \right)=\Re \left[\chi _{i}\left(\omega \right) \right],{\rm \; }i=1,\ldots ,N. 
\end{equation} 

 We can use the right polar decomposition of the fiber transfer matrix ${\bf H}(\omega )$ \cite{Huttner}
\begin{equation} \label{eq:polarDecomposition} 
{\bf H}(\omega )={\bf U}(\omega ){\bf P}(\omega ), 
\end{equation} 
where ${\bf U}(\omega )$ is a unitary matrix and ${\bf P}(\omega )$ is a positive-semidefinite Hermitian matrix.

Multiplying  ${\bf H}(\omega )$ by its adjoint yields
\begin{equation} \label{18)} 
{\bf H}(\omega )^{\dag } {\bf H}(\omega )={\bf P}(\omega )^{\dag } {\bf P}(\omega )={\bf P}(\omega )^2. 
\end{equation} 

Since ${\bf P}(\omega )$ is a Hermitian positive-semidefinite matrix, its eigenvalues are real and its eigenvectors  corresponding to different eigenvalues are orthogonal to each other. 

Using the spectral decomposition of ${\bf P}(\omega )^2$ in terms of its eigenvalues $\exp\left[{-a_{k} \left(\omega \right)z}\right]$ and eigenvectors ${\left| \upsilon _{k} \left(\omega \right) \right\rangle} $, we obtain
\begin{equation}  
{\bf P}(\omega )^2=\sum _{k=1}^{N}e^{-a_{k} \left(\omega \right)z} {\left| \upsilon _{k} \left(\omega \right) \right\rangle} \langle \upsilon _{k} \left(\omega \right)|,
\label{eq:MDLoperator}
\end{equation} 
where $z$ represents the fiber length.

We refer to $a_{k} \left(\omega \right),{\left| \upsilon _{k} \left(\omega \right) \right\rangle} $ as the {\em  principal attenuation coefficients} and the {\em principal attenuation modes}, respectively. The latter are pairwise orthogonal. 

We define MDL as the ratio of the largest eigenvalue to the smallest eigenvalue of ${\bf P}(\omega )^2$
\begin{equation} \label{16)} 
{\rm MDL:}=\frac{\max \left\{e^{-a_{k} \left(\omega \right)z} \right\}_{k=1}^N}{\min \left\{e^{-a_{k} \left(\omega \right)z}  \right\}_{k=1}^N} . 
\end{equation} 

This is identical to taking the SVD of the fiber transfer matrix ${\bf H}(\omega )$ and defining  MDL as the ratio of the squares of the largest and the smallest singular values \cite{Fontaine:OFC}.

From (\ref{jonesToop:eq}), the projection operators ${\left| \upsilon _{k} \left(\omega \right) \right\rangle} \langle \upsilon _{k} \left(\omega \right)|$ can be decomposed into the basis of the identity matrix and the Gell-Mann matrices  \cite{Roudas:PJ17}
\begin{equation}  
{\left| \upsilon _{k} \left(\omega \right) \right\rangle} \langle \upsilon _{k} \left(\omega \right)|=\frac{1}{N} {\bf I}+\frac{1}{2C_{N} } \hat{\upsilon }_{k} \left(\omega \right)\cdot\boldsymbol{\mathrm{\Lambda }} . 
\label{eq:MDLmodeProjectionOperator}
\end{equation} 

Therefore, combining (\ref{eq:MDLoperator}) with (\ref{eq:MDLmodeProjectionOperator}) yields \cite{Andrusier}
\begin{equation}  
{\bf P}(\omega )^2=\alpha_{0} (\omega )\left[{\bf I}+\frac{1}{2C_{N} } \vec{\Gamma }(\omega )\cdot\boldsymbol{\mathrm{\Lambda }}\right] , 
\label{eq:MDLoperatorFinalExpansion}
\end{equation} 
where we defined the mean attenuation as the arithmetic mean of the eigenvalues of ${\bf P}(\omega )^2$
\begin{equation} \label{22)} 
\alpha _{0} (\omega ):=\frac{1}{N} \sum_{k=1}^{N}e^{-a_{k} \left(\omega \right)z}  , 
\end{equation} 
and the MDL vector 
\begin{equation} \label{23)} 
\vec{\Gamma }\left(\omega \right):=\frac{1}{\alpha _{0}(\omega )} \sum_{k=1}^{N}e^{-a_{k} \left(\omega \right)z}  \hat{\upsilon }_{k} \left(\omega \right). 
\end{equation}

We can determine $\alpha_{0},\vec{\Gamma }\left(\omega \right)$ by sending CW light into the fiber under test and measuring the average output power
\begin{equation} \label{24)} 
\bar{P}(z)=\bar{P}(0)\langle s|{\bf H}(\omega )^{\dag } {\bf H}(\omega ){|s \rangle}  , 
\end{equation} 
or, equivalently, using (\ref{eq:MDLoperatorFinalExpansion}), we can define the attenuation $\alpha (z)$ as
\begin{equation} 
\alpha (z): = \frac{\bar{P}(z)}{\bar{P}(0)}=\alpha _{0} (\omega )\left[1+\frac{1}{2C_{N}^2 } \vec{\Gamma }(\omega )\cdot \hat{s}\right]. 
\label{eq: averagePowerFinal} 
\end{equation} 

Due to the similarity of (\ref{eq: averagePowerFinal}) to (\ref{eq:MGD}), $\alpha_{0} (\omega ),\vec{\Gamma }(\omega )$ can be estimated using a  similar process to the one described in Sec. \ref{sec:MDvectorEstimation}. Now, we simply measure the average output power instead of the time-of-flight of pulses. An alternative method for estimating MDL based on power measurements using a direct-detection receiver is described in \cite{Choutagunta}.

Once $\alpha _{0} (\omega ),\vec{\Gamma }(\omega )$ are known, we can construct ${\bf H}(\omega )^{\dag }{\bf H}(\omega )$ and calculate
\begin{equation} \label{26)} 
{\bf P}(\omega )=\sqrt{{\bf H}(\omega )^{\dag } {\bf H}(\omega )} . 
\end{equation} 

We can optically compensate for the optical fiber's MDL at the fiber input. The  transfer matrix of the MDL-equalized optical fiber  will then be 
\begin{equation} \label{27)} 
{\bf U}(\omega )={\bf H}(\omega ){\bf P}(\omega )^{-1} . 
\end{equation} 

Since ${\bf U}(\omega )$ is unitary,  the input group delay operator $i{\bf U}^{\dag } (\omega ){\bf U}_{\omega } (\omega )$ can be expressed as a linear combination of the identity matrix and the Gell-Mann matrices as in (\ref{eq:MDvectorDefinition})
\begin{equation} \label{28)} 
i{\bf U}^{\dag } (\omega ){\bf U}_{\omega } (\omega )=\tau _{0}^{'}(\omega ) {\bf I}+\frac{1}{2C_{N} } \vec{\tau }_{s}^{'} \left(\omega \right)\cdot \boldsymbol{\mathrm{\Lambda }} . 
\end{equation} 
Using the  process described in Sec. \ref{sec:MDvectorEstimation}, we can estimate $\tau _{0}^{'}(\omega ) ,\vec{\tau }_{s}^{'} \left(\omega \right)$. 

%Notice that $\tau _{0}^{'} ,\vec{\tau }_{s}^{'} \left(\omega \right)$ are not equal to $\tau _{0}^{} ,\vec{\tau }_{s}^{} \left(\omega \right)$ computed in the absence of MDL.

By differentiating (\ref{eq:polarDecomposition}) with respect to the angular frequency $\omega$, we obtain
\begin{multline*}
 \label{29)} 
i{\bf H}(\omega )^{-1} {\bf H}_{\omega } (\omega )={\bf P}^{-1} (\omega )\left[i{\bf U}^{\dag } (\omega ){\bf U}_{\omega } (\omega )\right]{\bf P}(\omega )\\+i{\bf P}^{-1} (\omega ){\bf P}_{\omega } (\omega ). 
\end{multline*} 

Since we have already calculated ${\bf P}(\omega ),i{\bf U}^{\dag } (\omega ){\bf U}_{\omega } (\omega ),$ we can calculate $i{\bf H}(\omega )^{-1} {\bf H}_{\omega } (\omega )$ and its eigenvalues and eigenvectors.

In summary, in the joint presence of MD and MDL, we  follow a divide-and-conquer approach, i.e., we conduct the experiment into two successive phases: (i) Initially, we perform MDL characterization  exclusively, followed by optical compensation of MDL;  (ii) Then, we measure the MD vector of the compensated fiber transfer matrix. At each stage, we perform sequential measurements for the various components of the MDL and MD vectors at different frequencies, following identical procedures, launching the same set of quasi-orthogonal vectors each time. Finally, we combine the experimental results to create the input group delay operator and calculate the input PMs and the corresponding DMGDs of the optical fiber.

\subsection{Practical considerations \label{Practical_considerations}}
\subsubsection{Modal crosstalk}

\noindent The mode-dependent signal delay method is vulnerable to errors in the mode converter settings. In this subsection, we aim to assess the impact of  modal crosstalk induced by erroneous mode converter settings. A back-of-the-envelope calculation, for the case of SMFs ($N=2$), indicates that this type of crosstalk can be the most significant limiting factor of the mode-dependent signal delay method. However, it can also be minimized by using quasi-orthogonal vectors in Stokes space as launch modes.

For MD characterization of FMFs and MMFs, mode selection at the fiber input can be accomplished, for instance,  by using a Liquid Crystal on Silicon (LCOS)-based spatial light modulator (SLM) as a mode converter \cite{Hoyningen}. These devices can be used to implement reconfigurable phase masks by varying the voltage of their pixels. Their main drawback is that they exhibit high crosstalk between certain pairs of modes \cite{Hoyningen}. The optimization of phase masks using simulated annealing can reduce the crosstalk introduced by the SLM \cite{Carpenter}. Nevertheless, any residual crosstalk in the transfer matrix of the mode converter may influence the accuracy of the measurement of the fiber input MD vector. 

Detailed physical modeling of the operation of the SLM is out of the scope of this paper. We consider an abstract model instead. Assume that we launch the perturbed states $\hat{s}_i^{'}$ instead of the intended states $\hat{s}_i$, due to  errors in the SLM settings. Then, we obtain erroneous group delay measurements that affect the entries of the DMGD vector ${\boldsymbol{\mathrm{T}}}_{\mathrm{g}}$ in (\ref{eq:dmgds}). The elements of the modified DMGD vector ${\bf T}_{g}^{'}$ are given by
\beq
{\bf T}_{g}^{'}={\bf S'}\vec{\tau }_{s},
\eeq 
where we defined the perturbed coefficient matrix
\begin{equation}
{\boldsymbol{\mathrm{S}}^{'}}  \df  \left[{\hat{s}}_1^{'},\dots {,\hat{s}}_{\ N^2-1}^{'}\right]^{{T}}.
\label{eq:perturbedCoefficientMatrix}
\end{equation}

Assuming that we are unaware of the errors in the SLM settings, we reconstruct the MD vector using the matrix inversion (\ref{eq:MDvectorCalculation})
\[\vec{\tau }_{s}^{'} ={\bf S}^{-1} {\bf T}_{g}^{'} .\] 

The computed MD vector $\vec{\tau }_{s}^{'} $ is offset from its nominal value $\vec{\tau }_{s}$ by
\beq\delta \vec{\tau }_{s}^{} =\vec{\tau }_{s}^{'} -\vec{\tau }_{s}^{} ={\bf S}^{-1} \delta {\bf S}\vec{\tau }_{s},
\label{eq:offsetMDvector}
\eeq 
where we defined the error matrix $\delta {\bf S}$
\beq
\delta {\bf S}:={\bf S}'-{\bf S}.
\label{eq:ds}
\eeq

We want to estimate the error in (\ref{eq:offsetMDvector}). For this purpose, we will use the following inequality for square matrices \cite[Theorem 2.10]{Suli}
\beq
\left\| {\bf A} {\bf B}\right\|\leq\left\| {\bf A}\right\|\left\|  {\bf B}\right\|,
\eeq
where $\left\| {\bf X}\right\|$ denotes the spectral norm of a matrix ${\bf X}$, which is the maximum singular value of ${\bf X}$.  

Taking the spectral norm of both sides of (\ref{eq:offsetMDvector}) yields
\beq\left\| \delta \vec{\tau }_{s} \right\| \le \left\| {\bf S}^{-1} \right\| {\rm \; }\left\| \delta {\bf S}\right\| {\rm \; }\left\| \vec{\tau }_{s} \right\|, \eeq 
or, equivalently, %\cite[Theorem 2.11]{Suli},
\beq\frac{\left\| \delta \vec{\tau }_{s} \right\| }{\left\| \vec{\tau }_{s} \right\| } \le \left\| {\bf S}^{-1} \right\| {\rm \; }\left\| \delta {\bf S}\right\| =\kappa \left({\bf S}\right){\rm \; }\frac{\left\| \delta {\bf S}\right\| }{\left\| {\bf S}^{} \right\| }, \eeq 
where $\kappa \left({\bf S}\right)$ is the condition number \cite[Definition 2.12]{Suli}
\beq\kappa \left({\bf S}\right):{\rm =\; }\left\| {\bf S}^{-1} \right\| {\rm \; }\left\| {\bf S}^{} \right\| .
\eeq

The condition number increases as the matrix gets closer to being singular. The choice of quasi-orthogonal Stokes vectors reduces $\kappa \left({\bf S}\right)$ but could affect $\left\| \delta {\bf S}\right\|$ as well.

As an illustrative example, consider the elementary case of a SMF  ($N=2$).  In this case, the mode converter can be substituted by a polarization controller and we can use the polarization-dependent signal delay method \cite{Nelson} for PMD characterization.

Assume that the intended launch vectors are linear horizontal, linear 45 deg, and right-circular SOPs, denoted by 
${{\left| s_{1}  \right\rangle} ={\left| {\rm e}_{{\rm x}}  \right\rangle} }, {{\left| s_{2}  \right\rangle} ={\left| {\rm e}_{{\rm 45}^{\circ } }  \right\rangle} }, {{\left| s_{3}  \right\rangle} ={\left| {\rm e}_{RC}  \right\rangle} }$, respectively. 

Due to the finite extinction ratio of the polarization controller, every time we attempt to launch a given SOP, we excite also its orthogonal SOP. The polarization crosstalk level $\varepsilon $  is defined as the ratio of the power launched at the orthogonal SOP and the nominal signal power. Since Jones vectors represent electric fields, the generated SOPs ${\left| s_{i}^{'}  \right\rangle} ,i=1,\ldots ,3$ contain perturbations of order $\sqrt{\varepsilon } $. Modeling the transfer matrix of the polarization controller as unitary, the actual launch vectors are  
\[\begin{array}{l} {{\left| s_{1}^{'}  \right\rangle} =\sqrt{1-\varepsilon } {\left| {\rm e}_{{\rm x}}  \right\rangle} +\sqrt{\varepsilon } {\left| {\rm e}_{{\rm y}}  \right\rangle} } \\ {{\left| s_{2}^{'}  \right\rangle} =\sqrt{1-\varepsilon } {\left| {\rm e}_{{\rm 45}^{\circ } }  \right\rangle} +\sqrt{\varepsilon } {\left| {\rm e}_{{\rm -45}^{\circ } }  \right\rangle} } \\ {{\left| s_{3}^{'}  \right\rangle} =\sqrt{1-\varepsilon } {\left| {\rm e}_{RC}  \right\rangle} +\sqrt{\varepsilon } {\left| {\rm e}_{LC}  \right\rangle} } \end{array}\] 
where  
${\left| {\rm e}_{{\rm y}}  \right\rangle} ,{\left| {\rm e}_{{\rm -45}^{\circ } }  \right\rangle}, {\left| {\rm e}_{LC}  \right\rangle}$, denote the linear vertical, linear -45 deg, and left-circular SOPs, respectively.

In the ideal case without polarization crosstalk, the  coefficient matrix ${\bf S}$ given by (\ref{eq:coefficientMatrix}) is just the identity matrix. In the presence of crosstalk, the actual coefficient matrix ${\bf S'}$ is
\beq
{\bf S'}=\left[\begin{array}{ccc} {1-2\varepsilon } & {2\sqrt{(1-\varepsilon )\varepsilon } }  & {0}\\ {2\sqrt{(1-\varepsilon )\varepsilon } } & {1-2\varepsilon } & {0} \\ {2\sqrt{(1-\varepsilon )\varepsilon } } & {0} &  {1-2\varepsilon } \end{array}\right]
\label{eq:sprime}
\eeq

Upon substituting (\ref{eq:sprime}) into (\ref{eq:ds}), we obtain
\beq
	\delta {\bf S'}=\left[\begin{array}{ccc} {-2\varepsilon } & {2\sqrt{(1-\varepsilon )\varepsilon } }  & {0}\\ {2\sqrt{(1-\varepsilon )\varepsilon } } & {-2\varepsilon } & {0} \\ {2\sqrt{(1-\varepsilon )\varepsilon } } & {0} &  {-2\varepsilon } \end{array}\right].
	\eeq

	Expanding in Taylor series with respect to the crosstalk level $\varepsilon $ yields
\beq 
	\delta {\bf S}=\left[\begin{array}{ccc} {O\left(\varepsilon \right)}  & {2\sqrt{{\rm }\varepsilon} +O\left(\varepsilon \right)}& {0} \\ {2\sqrt{{\rm }\varepsilon} +O\left(\varepsilon \right)} & {O\left(\varepsilon \right)} & {0} \\  {2\sqrt{{\rm }\varepsilon} +O\left(\varepsilon \right)} & {0} & {O\left(\varepsilon \right)} \end{array}
	\right].
\eeq

The spectral norm of the error matrix $\delta {\bf S }$ is
\beq
\left\| \delta {\bf S}\right\| =2\sqrt{2\varepsilon } +O(\varepsilon ).
\eeq 

The bound of the relative error in the MD vector is given by
\beq
\frac{\left\| \delta \vec{\tau }_{s} \right\| }{\left\| \vec{\tau }_{s} \right\| } \le \left\| {\bf S}^{-1} \right\| {\rm \; }\left\| \delta {\bf S}\right\| =2\sqrt{2\varepsilon } +O(\varepsilon ).\eeq
 
We observe that the relative error in the PMD vector is of order $\sqrt{\varepsilon } .$ It is crucial that we know the launched SOPs precisely, that is with ${\varepsilon }<-40$ dB, in order for the relative error in the assessment of the PMD vector ${\left\| \delta \vec{\tau }_{s} \right\| }/{\left\| \vec{\tau }_{s} \right\| }$ to be of the order of 1\%. For $N>2$, modal crosstalk is more severe. This places extremely stringent requirements on the accuracy with which we know the launch states into the fiber. In practice, this means that the transfer matrix of the mode converter is an inseparable part of the optical fiber transfer matrix and influences the measurements. All MD characterization methods essentially measure the joint transfer matrix of the  input/output spatial multiplexer/demultiplexer and the optical fiber \cite{Fontaine:OFC}.

\subsubsection{Numerical quadrature accuracy}
At the direct-detection receiver, the signal is sampled and the group delay $\tau_g$ is estimated by evaluating the integral (\ref{eq:tg}) numerically.  The computation can be carried out by using any numerical integration technique for equally-spaced subdivisions \cite{NR}. The accuracy of the integral depends on the sampling frequency and the vertical resolution of the real-time oscilloscope, as well as the pulse shape, the integration interval $T$, and  the particular quadrature rule. 

Assume that we launch ideal Gaussian pulses with half-width at the $1/e$ power point \cite{Agrawal} equal to $T_{0}$=10 ns. The integration time in (\ref{eq:MGDerror}) is $T=50$ ns. Using a low-end, real-time oscilloscope with 2 GHz bandwidth, 5 GSa/s sampling frequency, and 16 b vertical resolution, the relative error for detecting 0.1 ps group delays is less than 1\% when using the composite 3/8 Simpson's rule for numerical quadrature \cite{NR}. Therefore, it is not necessary to employ a fast  real-time oscilloscope or fs pulses in order to achieve sub-ps resolution.

\subsubsection{Thermal noise significance}
The reconstruction  of the MD vector is influenced by the thermal noise of the individual measurements. For $N\sim 100$, $N\sim 10^4$ sequential measurements are required to recover the $N\sim 10^4$ components of the MD vector. During the MD vector reconstruction, the variances of the noises of  individual measurements add up. The noise impact on the MD vector ends up being ~40 dB higher than the impact of the noise on each individual measurement.

Consider a thermal-noise limited direct-detection receiver with noise-equivalent power \cite{Agrawal} equal to 10 pW/$\sqrt{\mathrm Hz}$. An integration time $T=50$ ns corresponds to a digital filter with 100 MHz noise-equivalent bandwidth. We assume that the bandwidth of the photodiode is much larger than this value so that the thermal noise is essentially filtered digitally. For 10 mW received power, the rms noise per measurement given by (\ref{eq:NoiseVarianceFinalExpression}) is on the order of 0.1 ps. Even for 100-mode SDM MMF, the rms noise given by (\ref{eq:MDVectorLengthVariance}) is on the order of 10 ps.

\section{Optimization formalism \label{sec:optimization1}}
\noindent We seek to compute a set of $N^2-1$ Stokes vectors ${\hat{s}}_i$, ${i}=1,{\dots}, N^2-1,$ that minimizes the variance of the MD vector.

\subsection{Cost function}
\noindent Neglecting $\sigma _{\delta T_{g} }^{2}$ in (\ref{eq:MDVectorLengthVariance}), since it is  dependent on the specific implementation of the direct-detection receiver, we adopt the squared Frobenius norm of ${\bf A}$ as a normalized cost function
\begin{equation}
\xi \df \left\| {\bf{A}} \right\|_F^2={\rm Tr}\left[{\bf A}{\bf A}^{T} \right].
\label{eq:costfunction}
\end{equation}  

We want to minimize $\xi$ subject to the constraint that the Stokes vectors in the matrix $\bf{S}$ must correspond to valid combinations of modes in the generalized Jones space.

An alternative form for the cost function (\ref{eq:costfunction}) is
\begin{equation}
\label{eq:objectiveFun}
\xi = \mytrace(\AA^{-1}), %= \mytrace( (\SSS\SSS^T)^{-1} ). 
\end{equation}
where we defined
\beq
\AA := {\SSS\SSS}^T.
\eeq 

Notice that the entries of $\AA$ are the inner products of the Stokes vectors, i.e., $\AA_{jk} = \sh_j\cdot \sh_k$. In other words, $\AA$ is the Gram matrix of the Stokes vectors $\sh_j$.

Since $\xi \geq N^2-1$, where the lower bound of $\xi = N^2-1$ occurs in the ideal case of orthonormal vectors, we define the penalty for choosing a set of $N^2-1$ non-orthonormal Stokes vectors ${\hat{s}}_i$, ${i}=1,{\dots}, N^2-1,$ as 
\beq
\delta\df\frac{\xi}{(N^2-1)}.
\label{eq:penalty}
\eeq
The latter quantity can be viewed as the noise amplification per degree of freedom.

\subsection{Gradient descent method}

\noindent  Assume that the Jones vectors $\ket{s_i}$, $i=1,\ldots,N^2-1$, and, thus, ${\bf S}$ and $\xi$, are functions of $n$ real parameters $p_{1} ,\ldots ,p_{n} .$ We can write the parameters in a concise form as a column vector  
\begin{equation} \label{1)} 
{\bf p}\df\left[p_{1} ,\ldots ,p_{n} \right]^{T} . 
\end{equation}

We define the gradient of $\xi$ as the column vector
\begin{equation} \label{2)} 
\nabla \xi \df\left[\frac{\partial \xi }{\partial p_{1} } ,\ldots ,\frac{\partial \xi }{\partial p_{n} } \right]^{T}.  
\end{equation}

% \noindent where 
% \[n=(2N-2)(N^{2} -1).\] 
% It is worth noting that the number of parameters grows as $n\sim N^{3} .$ 

The method of gradient descent \cite{Boyd} uses an iterative algorithm to calculate a minimum of the cost function $\xi$. Starting from a given point ${\bf p}^{(0)} $, it makes successive steps to points  ${\bf p}^{(k)} $ by moving opposite to the direction of the gradient, until it reaches a local minimum:
\begin{equation} \label{3)} 
{\bf p}^{(k+1)} ={\bf p}^{(k)} -\mu^{(k)} \nabla \xi \left[{\bf p}^{(k)}\right],  
\end{equation} 
where $\mu^{(k)} $ is a positive constant (adaptive step size) \cite{Boyd}. This iterative process is continued until the magnitude of the gradient falls below a certain threshold or until a maximum number of iterations is reached.

From (\ref{eq:costfunction}), the components of the gradient of  $\xi$ can be written as
\begin{equation} \frac{\partial \xi }{\partial p_{r} } =2{\rm Tr}\left[\frac{\partial {\bf A}}{\partial p_{r} } {\bf A}^{T} \right].
\label{eq:gradA}
\end{equation} 

One can prove the above expression by writing the trace expression as a scalar using index notation, taking the derivative with respect to $p_r$, and rewriting the final result in matrix form.

We also use the following matrix identity 
\begin{equation} \frac{\partial {\bf A}}{\partial p_{r} } =-{\bf A}\frac{\partial {\bf S}}{\partial p_{r} } {\bf A}.
\label{eq:matrixDerivativeIdentity}
\end{equation} 

The trace of a product of matrices is invariant under cyclic permutation of the matrices in the product \cite{Strang}. Based on this property, we can rewrite (\ref{eq:gradA}) after substituting (\ref{eq:matrixDerivativeIdentity}) as
\begin{equation}\frac{\partial \xi }{\partial p_{r} } =-2{\rm Tr}\left[\frac{\partial {\bf S}}{\partial p_{r} } {\bf B}\right],
\label{eq:derpi}
\end{equation}
where we defined the auxiliary matrix
\begin{equation}
{\bf B}\df{\bf A A}^{T} {\bf A}.
\end{equation}

Assume that only one Stokes vector of the set $\hat{s}_1,\dots ,\hat{s}_{N^2-1}$ is a function of the parameter $p_{r }$. Furthermore, assume that this is the $k$-th Stokes vector $\hat{s}_k$. After taking the trace in (\ref{eq:derpi}), only the product of the $k$-th row of the first matrix with the $k$-th column of the second matrix remains. We write in a shorthand manner
\begin{equation}\frac{\partial \xi }{\partial p_{r} } =-2\frac{\partial \hat{s}_{k}^{T} }{\partial p_{r } } {\bf B}_{k} ,
\label{eq:IRfinalDerivativeXi}
\end{equation}
where ${\bf B}_{k}$ denotes the $k$-th column of the matrix ${\bf B}$. 

%and we introduced the index notation
%\beq
%k \df\left\lfloor \frac{r}{2n-2} \right\rfloor +1. 
%\eeq

In addition, we need to satisfy the constraint that the Stokes vectors making up the matrix ${\bf S}$  should correspond to valid combinations of modes in the generalized Jones space. We defer the discussion about how to take this constraint into account until Sec. \ref{sec:algorithms}.

%%%%%%%%%%%%%%%%%%%%%%%%%%%%%%%%%%%%%%%%%%%%%%%%%%%%%%%%%%%%%%%%%%%%%%%%
%\subsection{Gradient and Hessian of Det [NEW]:: MOD}
%%%%%%%%%%%%%%%%%%%%%%%%%%%%%%%%%%%%%%%%%%%%%%%%%%%%%%%%%%%%%%%%%%%%%%%%

%\pagebreak 
\subsection{Alternative formulation using Jones vectors}

\noindent To speed up numerical optimization, we find that it is computationally advantageous to express the elements of the gradient in terms of Jones vectors instead of Stokes vectors. This avoids unnecessary transitioning between Jones and Stokes spaces and eliminates the need for Gell-Mann matrices. 

As a starting point, we first calculate the derivative of the Stokes vector $\hat{s}_k$ with respect to $\pp_r$ by differentiating (\ref{eq:StokesVectors})
\beq
\frac{\partial  \sh_{k}}{\partial \pp_r}
= 2 C_N \Re \left\{\left\langle  s_k \middle| \bLambda \middle| \frac{\partial s_k}{\partial \pp_r}  \right\rangle \right\}.
\label{diffStokJones:eq}
\eeq

Furthermore, we notice that matrix  $\BBB$ can be expressed as $\BBB ={\SSS}^T \AA^{-2} $, so we rewrite (\ref{eq:derpi}) as
\begin{equation}\frac{\partial \xi }{\partial p_{r} } =-2{\rm Tr}\left[\AA^{-2} \frac{\partial {\bf S}}{\partial p_{r} }  {\SSS}^T\right].
\label{eq:derpialt}
\end{equation}
In (\ref{eq:derpialt}), we invoked the cyclical property of the trace and we pre-multiplied ${\partial {\bf S}}/{\partial p_{r} }  $ with $\AA^{-2}$ and post-multiplied with ${\SSS}^T$.

Assume that only the $k$-th Stokes vector $\hat{s}_k$ is a function of the parameter $p_{r }$. It follows that we can rewrite the previous expression into the form
\begin{equation}\frac{\partial \xi}{\partial \pp_r} =  -2 \sum_j \AA_{jk}^{-2}\left( \sh_j\cdot\frac{\partial \sh_k}{\partial \pp_r}\right).
\label{eq:derpialt2}
\end{equation}

We can rewrite the term inside the parenthesis in (\ref{eq:derpialt2}) using (\ref{diffStokJones:eq}) and the property [A.13] in \cite{Gordon}
\begin{equation} \sh_j\cdot\frac{\partial \sh_k}{\partial \pp_r}=2 C_N \Re \left\{\left\langle  s_k  \middle|\left( \sh_j\cdot \bLambda \right) \middle| \frac{\partial s_k}{\partial \pp_r} \right\rangle \right\}.
\label{eq:dotProduct1}
\end{equation}

Using (\ref{jonesToop:eq}), we obtain
\begin{align}
\label{gradGMriddanceBis:eq}
\frac{\partial \xi}{\partial \pp_r}
%  &= - 4 C_N \Re \ \sum_{j}   \BBB_j^T \left\langle s_j \middle| \bLambda \middle|  \frac{\partial s_j}{\partial \pp_r} \right\rangle \notag \\
&= - 8 C_N^2 \Re \left[ \sum_{j}   \AA_{jk}^{-2}  \langle s_k | s_j \rangle \left\langle s_j  \middle|  \frac{\partial s_k}{\partial \pp_r} \right\rangle \right]\notag \\
& + \frac{8 C_N^2}{N} \Re \left[\sum_{j}   \AA_{jk}^{-2}  \left\langle s_k  \middle|  \frac{\partial s_k}{\partial \pp_r} \right\rangle \right].
\end{align}

Also recall  that  (\ref{JonesOpTostokesInnerProd:eq}) gives  
\beq
\label{JonesOpTostokesInnerProdBis:eq}
\AA_{jk} =\sh_j \cdot \sh_k
= 2C_N^2 \left[ | \langle s_j |  s_k \rangle |^2 - \frac{1}{N}  \right].
% - \frac{1}{N-1} \langle s_j | s_j \rangle \langle s_k | s_k \rangle.
\eeq

From (\ref{gradGMriddanceBis:eq}), (\ref{JonesOpTostokesInnerProdBis:eq}), it is apparent that  ${\partial \xi}/{\partial \pp_r}$  can be expressed in terms of Jones vectors exclusively. Furthermore, (\ref{gradGMriddanceBis:eq}) uses the  inverse of  $\AA$, which is symmetric and positive semi-definite, unlike (\ref{eq:IRfinalDerivativeXi}) that is a function of ${\bf A}$, the inverse of $\SSS$. The advantage of (\ref{gradGMriddanceBis:eq}) over (\ref{eq:IRfinalDerivativeXi}) is that inversion of positive semi-definite matrices can be done via  Cholesky decomposition \cite{Strang}, which is twice as fast as the LU decomposition used for general matrices. Finally,  we shall see later on, when we discuss  the projected gradient method, that due to the constraints $\langle s_j | s_j \rangle = 1$, the terms in (\ref{gradGMriddanceBis:eq}) not only simplify slightly but the entire second sum in (\ref{gradGMriddanceBis:eq}) can be omitted, since it is orthogonal to the constraint manifold (cf. Sec. \ref{sec:algorithms}).

\section{Optimization algorithms\label{sec:algorithms}}

\noindent Our aim is to find an ``almost orthogonal" matrix ${\bf S}$ that minimizes the cost function $\xi$. In other words, we want to compute a set of $N^{\mathrm{2}}\mathrm{-1}$ maximally-orthogonal Stokes vectors ${\hat{s}}_i\ $that correspond to feasible combinations of propagating modes. This section is devoted to the description of two different gradient descent algorithms  \cite{Boyd} that can accomplish this task.

\subsection{Hyperspherical coordinates and unconstrained gradient descent}
\noindent In the first algorithm, we parameterize the \textit{j}-th unit Jones vector $\left.\mathrm{|}s_j\right\rangle \ $ by using $2N\mathrm{-}\mathrm{2\ }$hyperspherical coordinates \cite{Roudas:PJ17} 
\begin{multline} 
{\left| s \right\rangle} \df \left[\cos (\phi _{1} ),\sin (\phi _{1} )\cos (\phi _{2} )e^{i\theta _{1} } ,\ldots ,\right. \\ \left. \sin (\phi _{1} )\cdots \sin (\phi _{N-2} )\sin (\phi _{N-1} )e^{i\theta _{N-1} } \right]^{T} . 
\end{multline} 

Furthermore, we define the parameter vector $\boldsymbol{\mathrm{p}}$ that contains the coordinates ${\phi }_{jv}$ and ${\theta }_{jv}\ $of all $N^{\mathrm{2}}\mathrm{-}\mathrm{1}$ Stokes vectors$\mathrm{.\ }$ Then, we perform unconstrained optimization in a real space of ${n=\left(N^{\mathrm{2}}\mathrm{-}\mathrm{1}\right)\times \left(2N\mathrm{-}\mathrm{2}\right)} $ dimensions using the method of gradient descent (\ref{3)}). 

For large $N$'s, when ${\boldsymbol{\mathrm{p}}}^{\left(0\right)}$ is selected randomly, the matrix ${\bf S}$ might become almost singular, and the cost function can initially assume very high values. In this case, we find that the convergence of the gradient descent method can be accelerated by first using the normalized gradient and a constant step size $\mu$ in (\ref{3)})
\beq
{\boldsymbol{\mathrm{p}}}^{\left(\mathrm{k+1}\right)}\mathrm{=}{\boldsymbol{\mathrm{p}}}^{\left(\mathrm{k}\right)}\mathrm{-}{\mu }\frac{\mathrm{\nabla }{\mathrm{\xi }}\left({\bf p}^{(k)}\right)}{\left\Vert \mathrm{\nabla }{\mathrm{\xi }}\left({\bf p}^{(k)}\right)\right\Vert},
\eeq 
for $k=0,1,2, \ldots$. 

When the value of the cost function is decreased below a certain threshold, we revert back to (\ref{3)}), and the adaptive step ${\mu }^{\left(\mathrm{k}\right)}$ is selected using the backtracking method \cite{Boyd}.

\subsection{Cartesian coordinates and projected gradient descent}

\noindent In the second algorithm,  we  parametrize the $j$-th Jones vector $|s_j \rangle = (s_{jv})_{v=1}^N \in  \mathbb{C}^{N}$ by the $2N$ real parameters $x_{jv}:= \Re(s_{jv})$ and $y_{jv}:=\Im(s_{jv})$. Therefore we have ${n:=(N^2-1)\times 2N}$ parameters, which we arrange into  one column vector ${\boldsymbol{\mathrm{p}}} \in \mathbb{R}^n = \mathbb{R}^{2N} \times \ldots \mathbb{R}^{2N}$ by   concatenating together the $N^2-1$ copies of $[x_{j1},  y_{j1}, \ldots, x_{jN}, y_{jN}]^T \in \mathbb{R}^{2N}$, $j=1, \ldots, N^2-1$.
%[Thus we coordinated our notation with John's.]

The modest price to pay in order to avoid using the numerically slower trigonometric functions in the parameterization, is the imposition of $N^2-1$ unit length constraints
\beq
\gamma_j(\ppp):=\langle s_j | s_j \rangle = \sum_{v=1}^N \left( x_{jv}^2 + y_{jv}^2 \right) = 1,
\eeq
where $j=1, \ldots,N^2-1.$

This is to say that the parameter vector ${\boldsymbol{\mathrm{p}}}$ is restricted to a $(N^2-1) \times (2N-1)$ dimensional manifold $\MM$ in $\mathbb{R}^n$ that is the Cartesian product of $N^2 -1$ unit spheres in $\mathbb{R}^{2N}$.
Any non-zero ${\boldsymbol{\mathrm{p}}} \in \mathbb{R}^n$ can be projected into $\MM$ by simply normalizing each block $[x_{j1},  y_{j1}, \ldots, x_{jN}, y_{jN}]^T$. We denote this projection by $\text{proj}({\boldsymbol{\mathrm{p}}})$.  

Furthermore, at every ${\boldsymbol{\mathrm{p}}} \in \MM$, we have $N^2-1$ vectors normal to $\MM$ given by the unit normals to the individual spheres,
\begin{equation}
\label{jNormal:eq}
\nn_j ( {\boldsymbol{\mathrm{p}}}) := \left[0, \ldots, 0,x_{j1},y_{j1}, \ldots, x_{jN},y_{jN}, 0, \ldots, 0 \right]^T,
\end{equation}
where $j=1, \ldots, N^2-1.$

Alternatively, these vectors are the normalized gradients of individual constraints:
\begin{multline}
\label{jGrad:eq}
\nn_j({\boldsymbol{\mathrm{p}}}) = \frac{1}{2}\nabla \gamma_j({\boldsymbol{\mathrm{p}}}) =   \frac{1}{2} \left[ \frac{\partial \gamma_j }{\partial \pp_r } \right]_{r=1}^n
\\ =   \left[ \Re \left\langle s_j  \middle|  \frac{\partial s_j}{\partial \pp_r} \right\rangle \right]_{r=1}^n.
\end{multline}

Given any vector $\vv \in \mathbb{R}^n$ attached at ${\boldsymbol{\mathrm{p}}} \in \MM$, we can readily decompose it into components that are tangent (parallel) and orthogonal (normal) to the constraint manifold $\MM$, $\vv = \vv_\parallel + \vv_\perp$, where
\begin{equation}\label{perpProj:eq}
\vv_\perp = \sum_{j=1}^{N^2-1} (\nn_j(\ppp)^T \vv) \  \nn_j(\ppp).
\end{equation}
% This procedure will be applied to the gradient of the objective 

Recall that the gradient $\nabla \xi({\boldsymbol{\mathrm{p}}})$ is the column vector of all the partial derivatives given by (\ref{gradGMriddanceBis:eq}) for $r=1, \ldots, n$. In view of (\ref{jGrad:eq}), the second sum in (\ref{gradGMriddanceBis:eq}) is  a linear combination of $\nn_j({\boldsymbol{\mathrm{p}}})$. Thus it is orthogonal to $\MM$ and can be omitted if we only need the  tangential component of $\nabla \xi({\boldsymbol{\mathrm{p}}})$. Hence, to reduce the computational burden, we replace the gradient by
\begin{equation*}
\left[ \nabla \xi({\boldsymbol{\mathrm{p}}}) \right]_\parallel =
- 8 C_N^2 \left[ \Re \ \sum_{j}   \AA_{jk}^{-2}  \langle s_k | s_j \rangle \left\langle s_j  \middle|  \frac{\partial s_k}{\partial \pp_r} \right\rangle \right]_{r=1, \parallel}^n ,  
\end{equation*}
where the subscript $\parallel$ indicates taking the tangential component of the quantity inside the parenthesis.

The  method of projected gradient descent starts from a given point $\ppp^{(0)} \in \MM$ and then makes successive steps to points $\ppp^{(k)} \in \MM$ by first moving in the direction opposite the tangential component of the gradient and  then projecting (renormalizing) to hop back onto $\MM$.  
The corresponding recursive formula is 
\begin{equation} \label{projGradStep:eq} 
\ppp^{(k+1)}
=\text{proj} \left\{ \ppp^{(k)} -{\mu }^{\left(\mathrm{k}\right)} \left[ \nabla \xi \left(\ppp^{(k)}\right)\right]_\parallel  \right\}
\end{equation} 
for $k=0,1,2, \ldots$ \cite{Boyd}.

To explore the sensitivity of both optimization algorithms to the choice of initial conditions, we perform several optimization runs with different random seeds. Given that the algorithms are computationally intensive, due to the high-dimensionality of the optimization space, individual runs are executed in parallel in a high-performance computing cluster.  

\begin{figure}[!ht]
	\centering
	\includegraphics[width=0.5\textwidth]{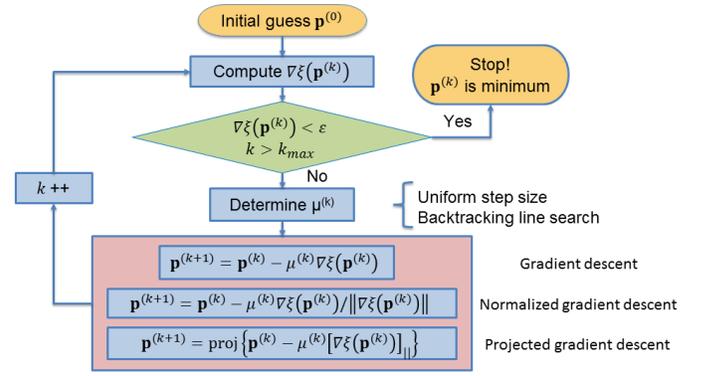}
	\caption{Flowchart of the gradient descent methods used for computing quasi-orthonormal vector sets in Stokes space.}
	\label{fig:flowchart}
\end{figure}

The flowchart shown in Fig. \ref{fig:flowchart} summarizes the steps of the numerical optimization process.

%\newpage

\section{Results and discussion \label{sec:results}}

\noindent In this section, we present the results of the numerical optimization algorithms described in Sec. \ref{sec:algorithms}. 

\begin{figure}
	\centering
	\begin{subfigure}[b]{0.4\textwidth}
		\includegraphics[width=1\linewidth]{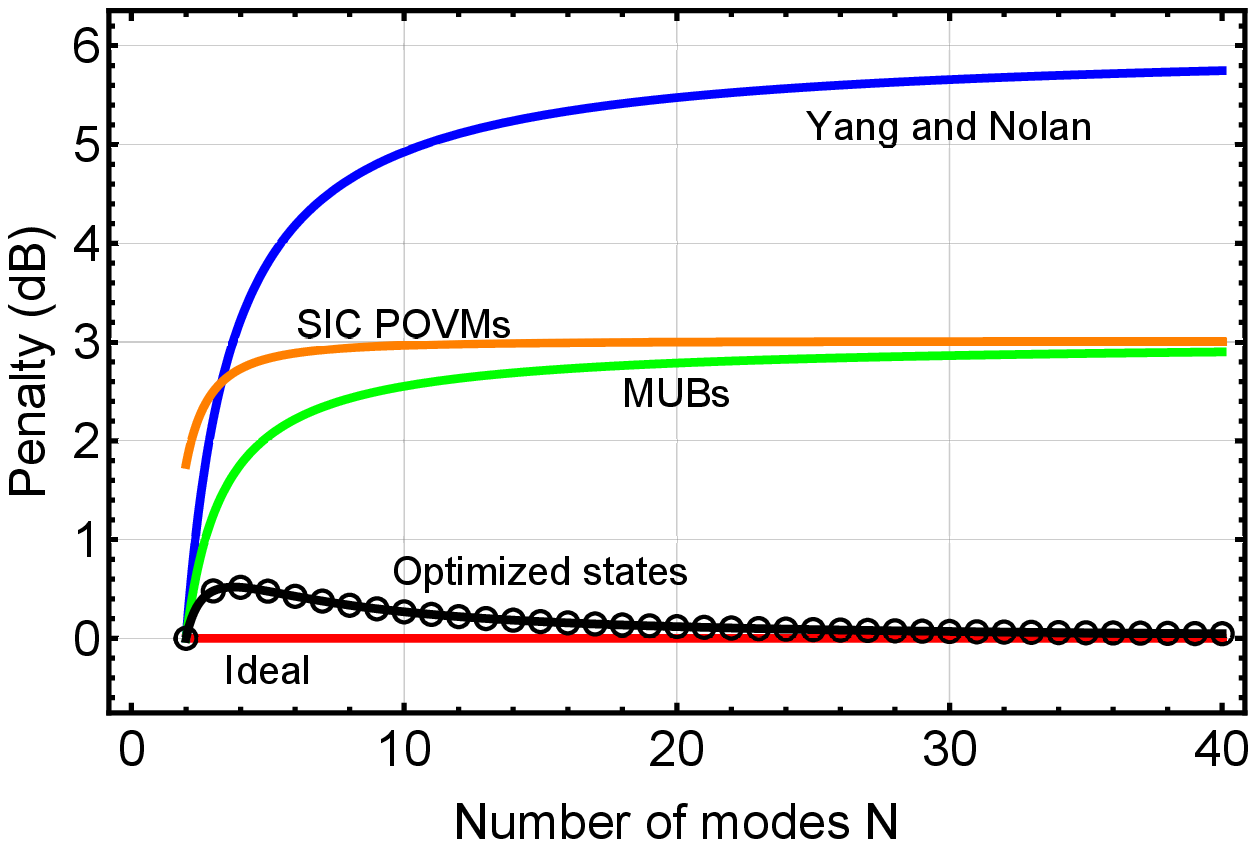}
		\caption{}
		\label{fig:Ng1} 
	\end{subfigure}
	
	\begin{subfigure}[b]{0.4\textwidth}
		\includegraphics[width=1\linewidth]{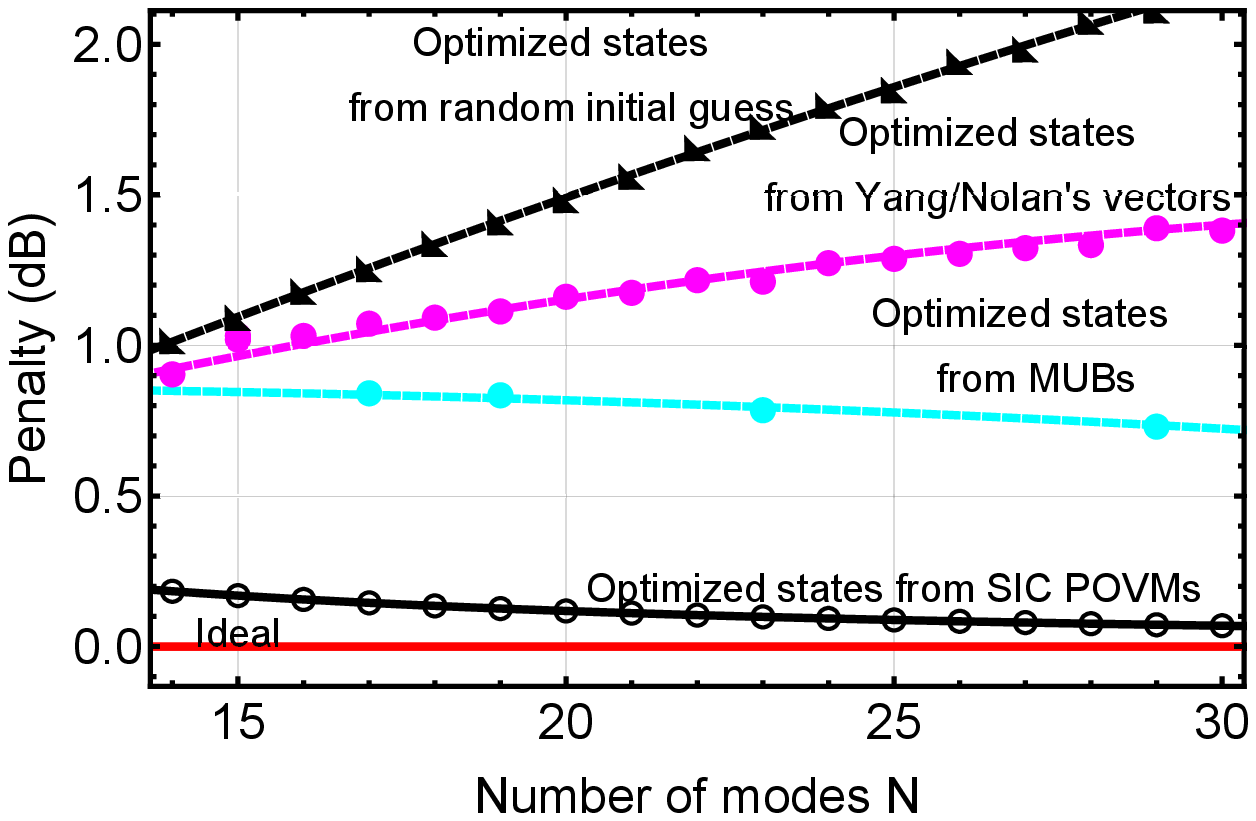}
		\caption{}
		\label{fig:Ng2}
	\end{subfigure}
	\caption[Numerical results]{(a) SNR penalty compared to the ideal case vs the number of modes for four different vector sets (Symbols: Blue line: Yang and Nolan's vectors \cite{Yang}; Green line: MUBs \cite{Bandyopadhyay}; Orange line: SIC POVMs \cite{Fuchs}; Red line: Orthonormal Stokes vectors; Circles: Numerical optimization using the algorithms in Sec. \protect\ref{sec:algorithms}. The optimization results were also validated by using the function FindMinimum in Mathematica \cite{Mathematica}, which is based on the Broyden-Fletcher-Goldfarb-Shanno (BFGS) quasi-Newton algorithm. Early results were presented in \cite{IR_CLEO_18}); (b) Sensitivity of the numerical optimization results to the initial conditions for $N=14-30$ (Symbols for different initial conditions: Triangles: Random initial guess; Magenta points: Yang and Nolan's vectors; Cyan points: MUBs; Open circles: SIC POVMs).}
\end{figure}

As an illustrative example, the optimal launch mode combinations for $N=4$ are given in Table \ref{tab:ss4}. Each mode is represented by a generalized Jones vector $\ket{s_i},i=1,\ldots, 15,$ and its decomposition in terms of the fiber eigenmodes $\ket{i},i=1,\ldots,4$, is listed. The angles among pairs of the corresponding generalized Stokes vectors vary in the interval 85$^\circ$-97$^\circ$, so the generalized Stokes vectors are approximately orthogonal. The value of the cost function for this vector set is $\xi = 16.9$, while for a truly orthogonal vector set it would be $\xi = 15$.  The SNR penalty is 0.517 dB. Penalties for the optimum vector sets for other values of $N$ are given in Fig.  \ref{fig:Ng1}.

Before we proceed, it would be instructive to explain how to use  the results of Table \ref{tab:ss4}. Consider the case of an FMF supporting the LP$_{01}$ and LP$_{11}$ mode groups. If there is weak coupling between the FMF mode groups, one can choose to characterize the modal dispersion of the LP$_{01}$ and LP$_{11}$ mode groups separately. In this case, the mode-dependent signal delay method  can be used first to determine the MD vector of the LP$_{01}$ mode group. This requires launching three different combinations of the x- and y-polarizations of the LP$_{01}$ mode. For this purpose, we can select three arbitrary orthonormal vectors in the conventional 3D Stokes space, e.g., the linear horizontal, linear 45 deg, and right-circular SOPs. Then,  the mode-dependent signal delay method can be applied once more to determine the MD vector of the LP$_{11}$ mode group. It is well known that the LP$_{11}$ mode group is composed of four spatial and polarization modes, i.e.,  the LP$_{11,o}$ and LP$_{11,e}$ modes, each in two orthogonal polarization configurations. The  mode-dependent signal delay method for $N=4$ requires launching 15 launch mode combinations. Launching the 15 launch mode combinations $\ket{s_i},i=1,\ldots, 15,$ shown in Table \ref{tab:ss4} will yield the smallest possible error in the measurement of the MD vector. The fiber eigenmodes $\ket{i},i=1,\ldots,4$, are the constituents of the LP$_{11}$ mode group. The  complex coefficients in  Table \ref{tab:ss4} represent the complex excitations of the phasors of the electric fields of these modes. 

      \begin{table}
	\centering
	\caption{Optimal vector set for $N$=4}
	% \begin{tabular}{|c|c|c{3cm}|c{3cm}|c| }
	%   \begin{tabular}{|c|c|c{1cm}|c{1cm}|c{1cm}|}
	\begin{tabular}{|c|c|c|c|c|} 
		% \hline\hline \multirow{2}{0.8cm}{Vector set} & \multicolumn{4}{c|}{\cellcolor{pink!25}Eigenmodes} \\
		\hline\hline \multirow{2}{0.8cm}{\centering Vector set} & \multicolumn{4}{c|}{\cellcolor{pink!25}Eigenmodes} \\
		% \cline{2-5} % does not work well 
		% \hhline{~|*4{-}}
		\hhline{~|----}
		% \mbox{} &  \cellcolor{pink!25} & \cellcolor{pink!25} & \cellcolor{pink!25} & \cellcolor{pink!25} \\
		\mbox{} & \cellcolor{pink!25}$\ket{1}$ & \cellcolor{pink!25}$\ket{2}$ & \cellcolor{pink!25}$\ket{3}$ & \cellcolor{pink!25}$\ket{4}$ \\
		% \mbox{} &   \cellcolor{pink!25} & \cellcolor{pink!25} & \cellcolor{pink!25} & \cellcolor{pink!25} \\
		% & \cellcolor{pink!25} & \cellcolor{pink!25} & \cellcolor{pink!25} & \cellcolor{pink!25} \\
		
		\hline
		\cellcolor{blue!25}$\ket{s_1}$ & 0.18 & $0.52-0.5 i$ & $-0.07+0.53 i$ & $-0.4-0.03 i$ \\	\hline
		\cellcolor{blue!25}$\ket{s_2}$ & 0.14 & $0.39+0.57 i$ & $-0.06-0.57 i$ & $-0.04+0.42 i$ \\	\hline
		\cellcolor{blue!25}$\ket{s_3}$ & 0.18 & $-0.68+0.35 i$ & $0.02 +0.43 i$ & $0.44 +0.08 i$ \\	\hline
		\cellcolor{blue!25}$\ket{s_4}$ & 0.23 & $-0.51-0.57 i$ & $-0.01-0.45 i$ & $-0.01-0.38 i$ \\	\hline
		\cellcolor{blue!25}$\ket{s_5}$ & 0.39 & $-0.24-0.02 i$ & $-0.58+0.43 i$ & $0.04-0.51 i$ \\	\hline
		\cellcolor{blue!25}$\ket{s_6}$ & 0.36 & $-0.05-0.26 i$ & $0.51-0.51 i$ & $-0.52+0.04 i$ \\	\hline
		\cellcolor{blue!25}$\ket{s_7}$ & 0.34 & $0.11-0.02 i$ & $-0.64+0.47 i$ & $0.09+0.49 i$ \\	\hline
		\cellcolor{blue!25}$\ket{s_8}$ & 0.34 & $0.16 i$ & $0.52-0.59 i$ & $0.48-0.07 i$ \\	\hline
		\cellcolor{blue!25}$\ket{s_9}$ & 0.45 & $-0.09+0.32 i$ & $-0.04-0.32 i$ & $-0.41-0.64 i$ \\	\hline
		\cellcolor{blue!25}$\ket{s_{10}}$ & 0.44 & $-0.46-0.13 i$ & $-0.03+0.18 i$ & $-0.62+0.4 i$ \\	\hline
		\cellcolor{blue!25}$\ket{s_{11}}$ & 0.42 & $-0.06-0.41 i$ & $-0.04-0.16 i$ & $0.52+0.59 i$ \\	\hline
		\cellcolor{blue!25}$\ket{s_{12}}$ & 0.42 & $0.36-0.02 i$ & $-0.03+0.14 i$ & $0.6-0.55 i$ \\	\hline
		\cellcolor{blue!25}$\ket{s_{13}}$ & 0.66 & $-0.39+0.31 i$ & $-0.31-0.38 i$ & $0.23+0.13 i$ \\	\hline
		\cellcolor{blue!25}$\ket{s_{14}}$ & 0.69 & $-0.45-0.34 i$ & $0.29 +0.24 i$ & $0.16-0.2 i$ \\	\hline
		\cellcolor{blue!25}$\ket{s_{15}}$ & 0.69 & $0.18-0.46 i$ & $-0.41-0.29 i$ & $-0.12-0.12 i$ \\	\hline
		\hline
		\multicolumn{5}{p{\columnwidth}}{Note: Jones vectors $\ket{s_i},i=1,\ldots, 15,$ corresponding to the optimal vector set for $N$=4. In each column we list their weights as a function of the fiber eigenmodes $\ket{i},i=1,\ldots,4.$}%\end{tablenotes}
		
	\end{tabular}
	\label{tab:ss4}
	
\end{table}

Next, it is shown that the computed optimal Stokes vector sets yield much better performance than previously proposed vector sets. Fig.  \ref{fig:Ng1} shows plots of the SNR penalty $\delta$  as a function of the number of propagation modes $N$ in the optical fiber for various vector sets. The ideal, albeit infeasible, case of orthonormal vectors is shown by the horizontal red line. The results of the numerical optimization are represented by the black curve with circles.  Notice that the penalty is initially 0 dB for $N=2$, reaches a maximum value for $N=4$, and then falls monotonically to almost 0 dB for  $N=40$. The fact that the penalty is 0 dB for $N=2$ comes as no surprise: in this case, the whole surface of the Poincar\'{e} sphere is covered with valid states. Thus, there exists an infinity of orthonormal vector sets that can be used for the measurement of the MD vector in Stokes space. For larger values of $N$, it is impossible to find an orthonormal set of $N^2-1$ Stokes vectors. For instance, for $N=4$, we observe that there is 0.517 dB penalty with respect to the ideal case. By further increasing $N$, we observe a gradual reduction in penalty, reaching 0.046 dB for $N=40$. 

For comparison, we included in the same graph, three additional plots corresponding to vector sets proposed in prior literature in optical communications and quantum mechanics, namely  Yang and Nolan's vectors \cite{Yang}, vectors selected from MUBs \cite{Bandyopadhyay}, and SIC-POVM vectors \cite{Fuchs} in blue, green, and orange, respectively (see Appendix A for details). The main advantage of these three vector sets is that there are relatively simple analytical or numerical algorithms for the evaluation of their coordinates. In contrast, computing the optimal vector sets using the method of gradient descent is time consuming for large values of $N$. On the downside, Yang and Nolan's vectors, MUBs, and SIC-POVMs present much higher penalties than the optimal vector sets given by numerical optimization. Indicatively, we remark the following features: a) $N^2-1$  Stokes vectors from MUBs can be evaluated using various numerical algorithms only for values of $N$ that are prime numbers or powers of prime numbers \cite{Bandyopadhyay}. It is worth noting that this vector set can be used only for waveguides lacking cylindrical symmetry. For the optical fibers of interest, the number of modes  $N$ can never be a prime or a power of a prime. We refer to MUBs here just for completeness. The green line in Fig.  \ref{fig:Ng1} shows that there is an asymptotic penalty equal to 3 dB for large $N$'s, i.e., the noise in the estimate of the MD vector is amplified by a factor of 2.    b) SIC POVMs exhibit slightly worse performance with respect to MUB vectors for small values of $N$ but, as $N$ increases, the penalty asymptotically reaches a ceiling of 3 dB, as in the case of MUB vectors (orange line in Fig.  \ref{fig:Ng1}). The main advantage of this vector family, compared to MUBs, is that $N^2-1$ vector sets can be computed for all practical values of $N$.   c) Yang and Nolan's vectors are given by simple analytical formulas, in contrast to the previous vector families. Nevertheless, they present worse performance than all prior vector sets, and the corresponding penalty asymptotically reaches 6 dB (blue line in Fig.  \ref{fig:Ng1}). 

In summary, the optimal vector sets provided by numerical optimization increase the SNR of the measurements  asymptotically by 3 dB for large values of $N$  compared to SIC POVMs and MUBs and by about 6 dB compared to Yang and Nolan's vectors. Therefore, we conclude that the performance of the mode-dependent signal delay can be dramatically improved by using the optimal vector sets provided by the gradient descent method. One can contrast this finding with the claim in Yang and Nolan's paper \cite{Yang} that the results of the mode-dependent signal delay method do not depend on the choice of launch state vectors, as long as the latter are linearly independent. This is true only in the absence of receiver noise.

\begin{figure}
	\centering
	\begin{subfigure}[b]{0.3\textwidth}
		\includegraphics[width=1\linewidth]{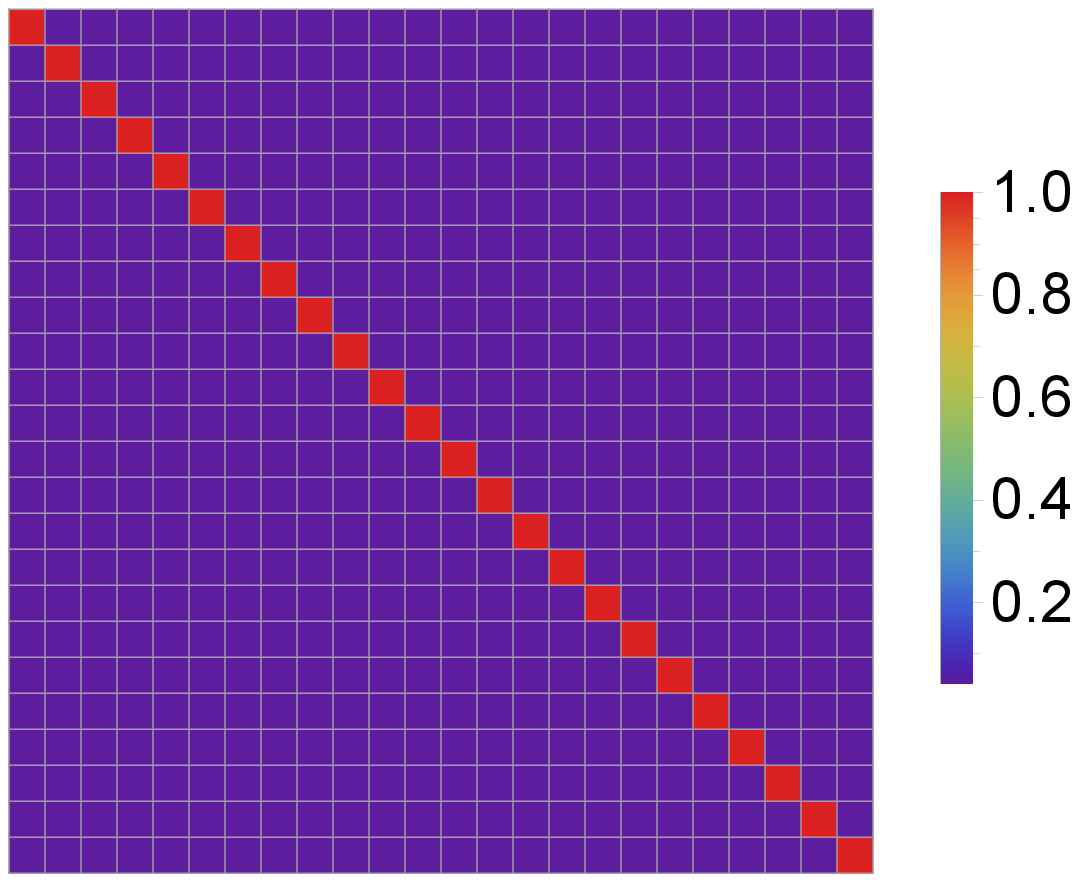}
		\caption{}
		\label{fig:plotCovarianceMatrixSICPOVMs}
	\end{subfigure}

	\begin{subfigure}[b]{0.3\textwidth}
		\includegraphics[width=1\linewidth]{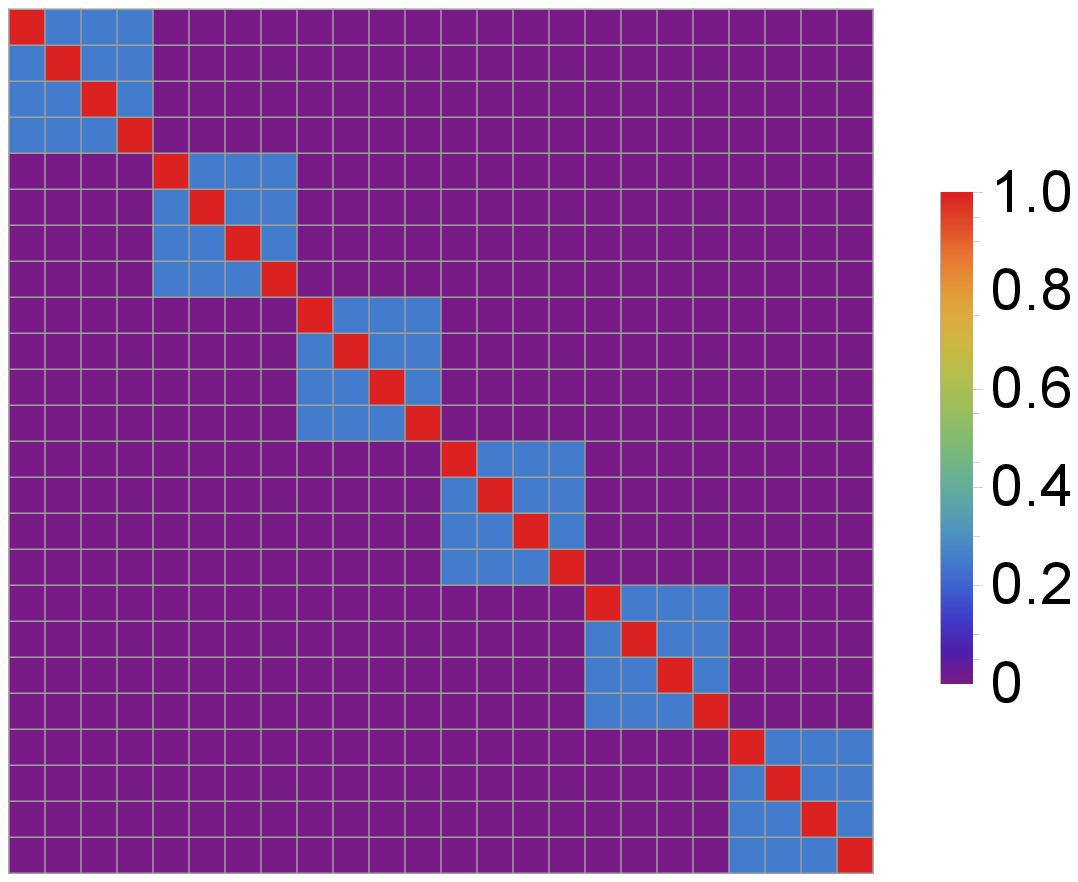}
		\caption{}
		\label{fig:plotCovarianceMatrixMUBs}
	\end{subfigure}
	
	\begin{subfigure}[b]{0.3\textwidth}
		\includegraphics[width=1\linewidth]{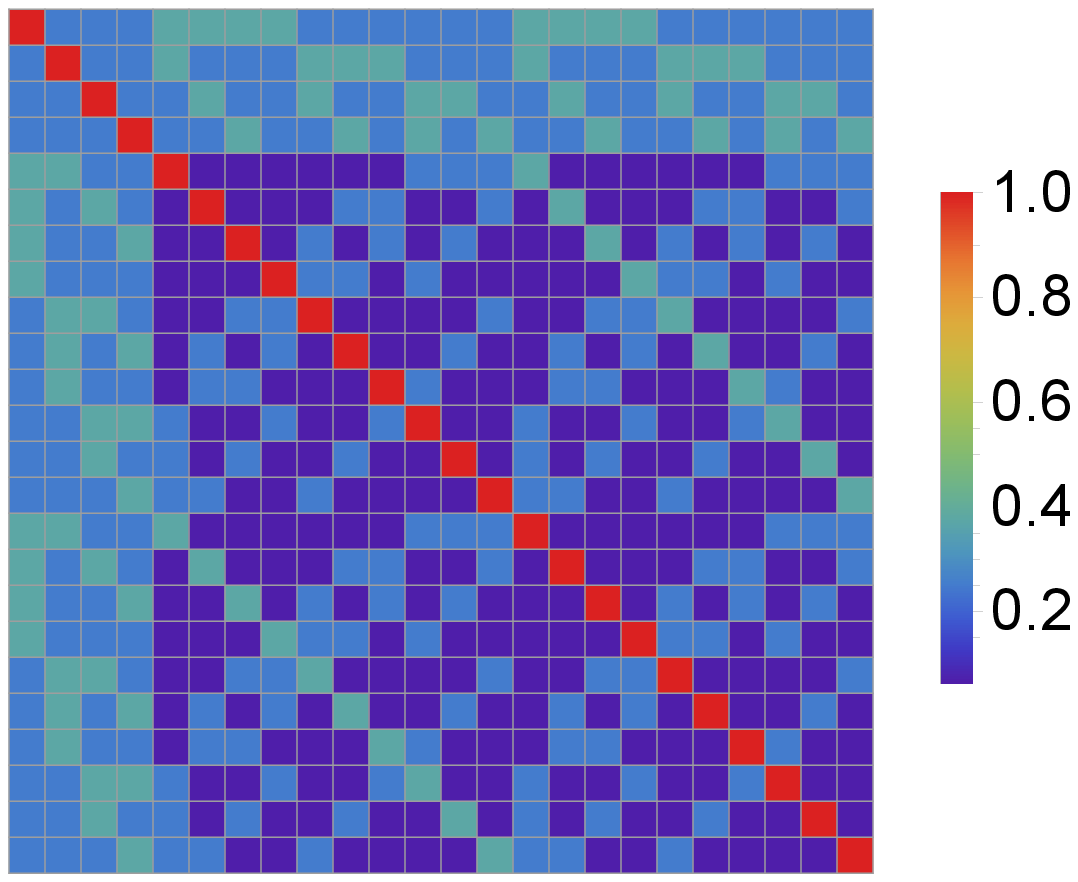}
		\caption{}
		\label{fig:CovarianceMatrixYang} 
	\end{subfigure}
	
	\label{fig:GramMatricesDensityPlot}
	\caption[Numerical results] {Density plots of the absolute value of the  Gram matrix $\AA$ of various  vector sets for \textit{N}=5: (a) SIC-POVMs \cite{Fuchs}; (b) MUBs \cite{Bandyopadhyay}; (c) Yang and Nolan's vectors \cite{Yang}.}
\end{figure}

 Despite their inferior performance compared to the optimal vectors, MUBs, SIC POVMs, and Yang and Nolan's vectors  are still useful as they can be used as starting points in order to accelerate the convergence of the gradient descent method. Fig. \ref{fig:Ng2} compares the results of the gradient descent method when the aforementioned vector sets are used as initial guesses for $N=15-30$. We observe  that the gradient descent method converges to different local minima for each vector family after 100,000 iterations. Interestingly, using SIC POVMs as initial guesses leads to the lowest penalties (black curve with circles) whereas the use of random initial vectors as  starting points leads to the worst performance (black curve with triangles). The use of MUB vectors and Yang and Nolan's vectors as starting points for the numerical optimization leads to intermediate penalty values in between the two back curves.   Indicative values of penalties are shown in Table 1.

\begin{table}
	\centering
	\caption{Numerical optimization results (Fig. \protect \ref{fig:Ng2}, $N$=30)}
	\begin{tabular}{ |c|c|c| } 
	\hline
	 \cellcolor{blue!25}  & \cellcolor{blue!25} & \cellcolor{blue!25}\\
	 \cellcolor{blue!25} Initial guess  & \cellcolor{blue!25} Initial penalty (dB) &  \cellcolor{blue!25} Final penalty (dB) \\  \cellcolor{blue!25} & \cellcolor{blue!25} & \cellcolor{blue!25} \\  \hline  & &\\ 
	SIC POVMs & 3.00 & 0.07 \\ \hline  & &\\
	MUBs & 2.86 & 0.73 \\ \hline & & \\
	Yang \& Nolan's vectors & 5.65 & 1.38 \\ \hline & & \\
	Random vectors & $>$50 & 2.2 \\ \hline 

\multicolumn{3}{p{\columnwidth}}{Note: Starting from different initial conditions, the numerical optimization reaches different local minima after 100,000 iterations. For instance, for $N=30$, using SIC POVMs in the mode-dependent signal delay method leads to a 3 dB penalty compared to the ideal case. However, using the gradient descent method with the SIC POVMs as an initial guess, we compute an optimal set of vectors that exhibits only 0.07 dB residual penalty at the end of the optimization process compared to the ideal case. Worse residual penalties are achieved by starting the optimization process using vectors from MUBs, Yang \& Nolan's vectors, and random vectors as initial conditions.}%\end{tablenotes}
\end{tabular}
\label{tab:ComparisonInitialConditions}
	
\end{table}

In retrospect, it is not surprising that the gradient descent method yields best results when the SIC POVM vectors are used as an initial guess: This must be attributed to the fact that SIC POVMs present maximum symmetry because they form a regular simplex in Stokes space and are equiangular, i.e., their pairwise inner products in Stokes space are the same. As explained in detail in  Appendix A, the pairwise inner product of two different SIC POVMs tends to zero for large values of $N$. This is illustrated by the almost diagonal Gram matrix for $N=5$ in Fig. \ref{fig:plotCovarianceMatrixSICPOVMs}.  For comparison, density plots of the Gram matrices for vectors from MUBs and Yang and Nolan's vectors are shown in Fig. \ref{fig:plotCovarianceMatrixMUBs} and Fig. \ref{fig:CovarianceMatrixYang}, respectively. The latter two density plots reveal a block diagonal and a block structure, respectively, indicative of much less symmetric vector configurations.

\begin{figure}[!ht]
	\centering
	\includegraphics[width=0.3\textwidth]{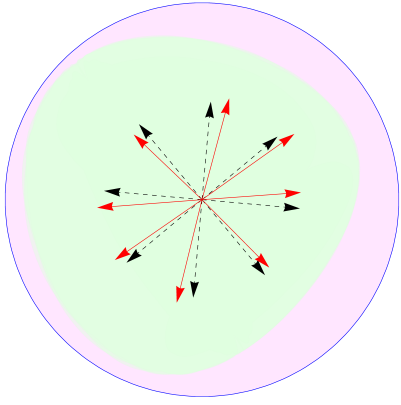}
	\caption{2D projections of various vector sets for \textit{N}=3 (Symbols: Red vectors: actual optimal vectors given by the numerical optimization of Sec. 2; Dashed black vectors: ideal orthonormal vectors best approximating the optimal vectors; Light green area: projection of the manifold of allowed states on the surface of the Poincar\'{e} sphere).}
	\label{fig:projections2D}
\end{figure}

Finally, we can catch a glimpse of the optimum set of Stokes vectors using a 2D projection (e.g., see Fig. \ref{fig:projections2D} for $N\mathrm{=3}$). We know that the $N^{\mathrm{2}}\mathrm{-}\mathrm{1}$ Stokes vectors should be ideally orthonormal. From the optimal vectors given by the numerical optimization procedure, we compute a set of  orthonormal vectors best approximating the optimal vectors. It is possible to project these orthonormal vectors onto a plane so that their projections have equal angular separations (dashed black vectors). Now we can superimpose on the same plane the projections of the actual optimal vectors given by the numerical optimization procedure (red vectors), as well as the projection of the manifold of allowed states on the surface of the Poincar\'{e} sphere (light green area) [5]. All vectors are bounded by the projection of the Poincar\'{e} sphere onto the plane (pink circular disk with unit radius).

\section{Summary}
\noindent In this article, we revised the mode-dependent signal delay method formalism for the characterization of SDM MMFs.  We analytically calculated the variance in the estimation of the length of the input MD vector due to receiver thermal noise. We showed that the mode-dependent signal delay method is versatile and can be applied to the estimation of the MDL vector, as well as the simultaneous measurement of the MD and MDL vectors.  We discussed various measurement errors other than these due to the thermal noise of the direct-detection receiver that occur during the characterization process. The latter part of the paper was devoted to the optimization of the launch states used in the mode-dependent signal delay method for the measurement of modal dispersion in SDM MMFs. The optimal sets of launch modes proposed here are universal, i.e., they are not limited to specific fiber types and can be used for SMFs, MMFs, and MCFs with strong and weak coupling.

As a final note, we stress that  all MD characterization methods inherently make measurement errors due to receiver noise,  modal crosstalk introduced during mode launch, and other implementation imperfections. Therefore, it is important to measure the modal dispersion of a fiber using various alternative methods and determine whether their results agree or not. Even if the mode-dependent signal delay method does not prevail as the method of choice for MD characterization, the set of optimal vectors proposed here can be used for measurements in the generalized Stokes space and data transmission using Stokes vector modulation. We anticipate that, since these vectors are quasi-orthonormal, they will give superior performance in a variety of problems involving measurements in Stokes space compared to  other vector sets proposed in the literature.

Optimum vector sets are available %from the authors upon request and 
online \cite{IR_Data_online}.

%\newpage

\appendices

% conference papers do not normally have an appendix

\section{Special vector sets\label{sec: Appendix}}
The goal of this Appendix is to derive useful analytical relationships for the cost function of special vector sets.

\subsection{Symmetric, informationally complete, positive operator valued measure (SIC-POVM) vectors \protect\cite{Fuchs}}

\noindent Consider the $N$-dimensional Jones space  $\mathbb{C}^{N} $.  The inner product of the $N^{2} $ unit SIC-POVM vectors $\left\{|\psi _{i} \rangle \right\}$  satisfies the condition
\begin{equation}
|\langle \psi _{i} |\psi _{j} \rangle |^{2} =\frac{1}{N+1} ,\quad \forall i\ne j
\label{eq:SICPOVMs}
\end{equation}

We recall that the dot product of Stokes vectors is related to the inner product of Jones vectors through (\ref{JonesOpTostokesInnerProd:eq}). By substituting (\ref{eq:SICPOVMs}) into (\ref{JonesOpTostokesInnerProd:eq}), we can compute the exact dot products of the SIC POVM vectors in generalized Stokes space
\begin{equation}
{\bf G}_{ij} =\hat{\psi }_{i} \cdot \hat{\psi }_{j} =\left\{\begin{array}{ccc} {1} & {} & {i=j} \\ {} & {} & {} \\ {-\frac{1}{N^{2} -1} } & {} & {i\ne j} \end{array}\right.
\label{eq:GramMatrixSICs} 
\end{equation}

A density plot of the Gram matrix ${\bf G}={\bf SS}^{T} $ for $N=5$ is shown in Fig. \ref{fig:plotCovarianceMatrixSICPOVMs}. Due to its simple structure, the cost function can be calculated analytically.

Namely, we can decompose ${\bf G}$ into a linear combination of two square $\left( N^2-1 \right) \cross \left( N^2-1 \right)$ matrices, the identity matrix ${\bf I }$ and the constant matrix ${\bf J}$ with all entries equal to unity
\begin{equation}
{\bf G }=\frac{N^{2} }{N^{2} -1} {\bf  I} -\frac{1}{N^{2} -1} {\bf J}
\label{eq:DecompositionGramSICs}
\end{equation} 

Matrices ${\bf I }$ and ${\bf J}$ commute so the eigenvalues of ${\bf G}$ are the corresponding combinations of the eigenvalues of these two matrices.

The identity matrix ${\bf  I}$ has a unit eigenvalue with multiplicity $N^{2} -1$, and the matrix ${\bf  J}$ has two eigenvalues, zero with multiplicity $N^{2} -2$ and $N^{2} -1$ with multiplicity 1. Therefore, the eigenvalues of ${\bf G }$ are
\begin{equation}
\lambda \left({\bf G}\right)=\left\{\underbrace{\frac{N^{2} }{N^{2} -1} ,\ldots \frac{N^{2} }{N^{2} -1} }_{N^{2} -2},\frac{1}{N^{2} -1} \right\}.
\end{equation}

Consequently, the eigenvalues of ${\bf G}^{-1} $ are
\begin{equation}\lambda \left({\bf G}^{-1} \right)=\left\{\underbrace{\frac{N^{2} -1}{N^{2} } ,\ldots ,\frac{N^{2} -1}{N^{2} } }_{N^{2} -2},N^{2} -1\right\}.
\label{eq:eigsInverseG}
\end{equation}

If ${\bf X}$ is a square $n \times n$ matrix, then the sum of the $n$ eigenvalues of ${\bf X}$ is the trace of ${\bf X}$ and the product of the $n$ eigenvalues is the determinant of ${\bf X}$.

Therefore, the  cost function can be analytically expressed as
\begin{equation}
\xi ={\rm Tr}\left({\bf G}^{-1} \right)=2\frac{\left(N^{2} -1\right)^{2} }{N^{2} }.
\end{equation} 

The penalty is given by
\begin{equation}
\delta =\frac{\xi }{N^{2} -1} =2\frac{\left(N^{2} -1\right)}{N^{2} }.
\end{equation}
 
For optical fibers supporting a large number of modes $N$, the penalty asymptotically reaches the limit
\begin{equation}
\mathop{\lim }\limits_{N\to \infty } \delta =2.
\end{equation}
We conclude that, for large $N$'s, there is roughly a 3 dB penalty compared to the ideal orthonormal states.

In addition, the volume of the parallelotope with edges equal to the Stokes vectors $\hat{s}_1,\dots,\hat{s}_{N^2-1}$ is given by the determinant of the Gram matrix $V =\sqrt{{\rm det}\left({\bf G}\right)}$. The last expression can be analytically evaluated from the eigenvalues in (\ref{eq:eigsInverseG})
\begin{equation}
V =\sqrt{{\rm det}\left({\bf G}\right)} =\frac{N^{N^{2} -2} }{\left(N^{2} -1\right)^{\frac{\left(N^{2} -1\right)}{2} } } 
\end{equation}

For optical fibers supporting a large number of modes $N$, the volume of the parallelotope asymptotically tends to zero
\beq
\mathop{\lim }\limits_{N\to \infty } V =0.
\eeq

\subsection{Vectors from mutually unbiased bases (MUBs) \protect\cite{Bandyopadhyay}}

Two distinct bases $\left\{|\psi _{i} \rangle \right\},\left\{|\phi _{j} \rangle \right\}$ are said to be mutually unbiased if $\hat{\psi }_{i} \cdot \hat{\phi }_{j} =0.$ Then, from  (\ref{JonesOpTostokesInnerProd:eq}), we obtain
\begin{equation}
|\langle \psi _{i} |\phi _{j} \rangle |^{2} =\frac{1}{N} ,\quad \forall i,j.
\label{eq:MUBsDotProduct}
\end{equation} 

There exist $N+1$ MUBs of $N$ vectors each when the number of modes \textit{N}  is a power of a prime \cite{Bandyopadhyay}. Here, we select launch states by picking groups of \textit{$N-1$ }vectors from each one of the \textit{$N+1\ $} MUBs.

From (\ref{JonesOpTostokesInnerProd:eq}) and (\ref{eq:MUBsDotProduct}), we can calculate  the elements of the covariance matrix ${\bf G}={\bf S S}^{T} $ without first calculating explicitly the MUB vectors.

It turns out that ${\bf G}={\bf S S}^{T} $is a $\left(N^2-1\right)\times \left(N^2-1\right)$ square matrix in block diagonal form
\begin{equation}
{\bf G}=\left[\begin{array}{cccc} {{\bf X}} & {0} & {0} & {0} \\ {0} & {{\bf X}} & {0} & {0} \\ {0} & {0} & {\ddots } & {0} \\ {0} & {0} & {0} & {{\bf X}} \end{array}\right],
\end{equation}
where ${\bf X}$ is a $\left(N-1\right)\times \left(N-1\right)$ submatrix given by
\begin{equation}
{\bf X}=\left[\begin{array}{cccc} {1} & {-\frac{1}{N-1} } & {\ldots } & {-\frac{1}{N-1} } \\ {-\frac{1}{N-1} } & {1} & {{\mathinner{\mkern2mu\raise1pt\hbox{.}\mkern2mu\raise4pt\hbox{.}\mkern2mu\raise7pt\hbox{.}\mkern1mu}} } & {-\frac{1}{N-1} } \\ {\vdots } & {{\mathinner{\mkern2mu\raise1pt\hbox{.}\mkern2mu\raise4pt\hbox{.}\mkern2mu\raise7pt\hbox{.}\mkern1mu}} } & {\ddots } & {-\frac{1}{N-1} } \\ {-\frac{1}{N-1} } & {\cdots } & {-\frac{1}{N-1} } & {1} \end{array}\right].
\end{equation}

This form is analogous to (\ref{eq:GramMatrixSICs}). Therefore, we can decompose ${\bf X}$ as in (\ref{eq:DecompositionGramSICs})
\begin{equation}
{\bf X}=\frac{N}{N-1}{\bf I}-\frac{1}{N-1} {\bf J},
\end{equation} 
where now ${\bf I},{\bf J}$ are square  $\left(N-1\right)\times \left(N-1\right)$ matrices.

Following the same methodology as in the preceding subsection, it is straightforward to show that the eigenvalues of ${\bf X}$ are
\begin{equation}
\lambda \left({\bf X}\right)=\left\{\underbrace{\frac{N}{N-1} ,\ldots ,\frac{N}{N-1} }_{N-2},\frac{1}{N-1} \right\}.
\end{equation}

For a block diagonal matrix
\begin{equation}
{\bf Y}=\left[\begin{array}{cccc} {{\bf Y}_{1} } & {0} & {\cdots } & {0} \\ {0} & {{\bf Y}_{2} } & {\cdots } & {0} \\ {\vdots } & {\vdots } & {\ddots } & {\vdots } \\ {0} & {0} & {\cdots } & {{\bf Y}_{n} } \end{array}\right],
\end{equation} 
the following properties hold
\begin{equation}
\begin{array}{c}{\det {\bf Y}=\prod_{i=1}^{n} \det {\bf Y_{i}} ,} \\ \\{{\rm Tr}{\bf Y}=\sum_{i=1}^{n}{\rm Tr}{\bf Y}_{i}.} \end{array}
\label{eq:MatrixProperties}
\end{equation}

Using (\ref{eq:MUBsDotProduct})-(\ref{eq:MatrixProperties}), the  cost function can be analytically expressed as
\begin{equation}
\xi ={\rm Tr}\left( {\bf G}^{-1} \right)=2\left(N^{2} -1\right)\frac{N-1}{N} .
\end{equation}
The penalty is given by
\begin{equation}
\delta =\frac{\xi }{N^{2} -1} =2\frac{N-1}{N} .
\end{equation}

For optical fibers supporting a large number of modes
\begin{equation}
\mathop{\lim }\limits_{N\to \infty } \delta =2
\end{equation} 
Asymptotically, there is roughly a 3 dB penalty compared to the ideal orthonormal states.

In addition, the volume can be analytically expressed as
\begin{equation}V =\sqrt{{\rm det}\left({\bf G}\right)} =\frac{N^{\frac{\left(N-2\right)\left(N+1\right)}{2} } }{\left(N-1\right)^{\frac{\left(N^{2} -1\right)}{2} } } .
\end{equation} 

For optical fibers supporting a large number of modes $N$, the volume of the parallelotope asymptotically tends to zero
\beq
\mathop{\lim }\limits_{N\to \infty } V =0.
\eeq

\subsection{Cost function for Yang and Nolan's vectors \protect\cite{Yang}}

\noindent Consider the fiber eigenmodes in Jones space ${\left| i \right\rangle} ,{\rm \; }i=1,\ldots ,N.$ Yang and Nolan's vectors are defined as \cite{Yang}
\beq\begin{array}{l} {{\left| x_{i}  \right\rangle} ={\left| i \right\rangle} {\rm \; \; \; }i=1,\ldots ,N-1} \\[10pt] {{\left| y_{ij}  \right\rangle} =\frac{{\left| i \right\rangle} {\rm +}{\left| j \right\rangle} }{2} {\rm \; \; \; }1\le i<j\le N} \\[10pt] {{\left| z_{ij}  \right\rangle} =\frac{{\left| i \right\rangle} {\rm +i}{\left| j \right\rangle} }{2} {\rm \; \; \; }1\le i<j\le N} \end{array}
\label{eq:YangVectors}
\eeq
In the last expression, we used different fonts to distinguish the imaginary number {\rm i} from the index $i$. Notice that there are $N^{2} -1$ vectors in total. 

%Their inner products in Jones space are
%\beq\begin{array}{l} {{\left\langle x_{i}  \mathrel{\left| \vphantom{x_{i}  x_{j} }\right.\kern-\nulldelimiterspace} x_{j}  \right\rangle} =\delta _{ij} } \\[10pt] {{\left\langle x_{i}  \mathrel{\left| \vphantom{x_{i}  y_{jk} }\right.\kern-\nulldelimiterspace} y_{jk}  \right\rangle} =\frac{\delta _{ij} +\delta _{ik} }{\sqrt{2} } } \\[10pt] {{\left\langle x_{i}  \mathrel{\left| \vphantom{x_{i}  z_{jk} }\right.\kern-\nulldelimiterspace} z_{jk}  \right\rangle} =\frac{\delta _{ij} +i\delta _{ik} }{\sqrt{2} } } \\[10pt] {{\left\langle y_{ij}  \mathrel{\left| \vphantom{y_{ij}  y_{k\ell } }\right.\kern-\nulldelimiterspace} y_{k\ell }  \right\rangle} =\frac{\delta _{ik} +\delta _{i\ell } +\delta _{jk} +\delta _{j\ell } }{2} } \\[10pt] {{\left\langle y_{ij}  \mathrel{\left| \vphantom{y_{ij}  z_{k\ell } }\right.\kern-\nulldelimiterspace} z_{k\ell }  \right\rangle} =\frac{\delta _{ik} +\delta _{jk} +i\left(\delta _{i\ell } +\delta _{j\ell } \right)}{2} } \\[10pt] {{\left\langle z_{ij}  \mathrel{\left| \vphantom{z_{ij}  z_{k\ell } }\right.\kern-\nulldelimiterspace} z_{k\ell }  \right\rangle} =\frac{\delta _{ik} +\delta _{j\ell } +i\left(\delta _{i\ell } -\delta _{jk} \right)}{2} } \end{array}
%\eeq

The squared norms of the inner products in Jones space are
\beq\begin{array}{l} {\left|{\left\langle x_{i}  \mathrel{\left| \vphantom{x_{i}  x_{j} }\right.\kern-\nulldelimiterspace} x_{j}  \right\rangle} \right|^{2} =\delta _{ij} } \\[10pt] {\left|{\left\langle x_{i}  \mathrel{\left| \vphantom{x_{i}  y_{jk} }\right.\kern-\nulldelimiterspace} y_{jk}  \right\rangle} \right|^{2} =\frac{\left(\delta _{ij} +\delta _{ik} \right)^{2} }{2} } \\[10pt] {\left|{\left\langle x_{i}  \mathrel{\left| \vphantom{x_{i}  z_{jk} }\right.\kern-\nulldelimiterspace} z_{jk}  \right\rangle} \right|^{2} =\frac{\delta _{ij} +\delta _{ik} }{2} } \\[10pt] {\left|{\left\langle y_{ij}  \mathrel{\left| \vphantom{y_{ij}  y_{k\ell } }\right.\kern-\nulldelimiterspace} y_{k\ell }  \right\rangle} \right|^{2} =\frac{\left(\delta _{ik} +\delta _{i\ell } +\delta _{jk} +\delta _{j\ell } \right)^{2} }{4} } \\[10pt] {\left|{\left\langle y_{ij}  \mathrel{\left| \vphantom{y_{ij}  z_{k\ell } }\right.\kern-\nulldelimiterspace} z_{k\ell }  \right\rangle} \right|^{2} =\frac{\left(\delta _{ik} +\delta _{jk} \right)^{2} +\left(\delta _{i\ell } +\delta _{j\ell } \right)^{2} }{4} } \\[10pt] {\left|{\left\langle z_{ij}  \mathrel{\left| \vphantom{z_{ij}  z_{k\ell } }\right.\kern-\nulldelimiterspace} z_{k\ell }  \right\rangle} \right|^{2} =\frac{\left(\delta _{ik} +\delta _{j\ell } \right)^{2} +\left(\delta _{i\ell } -\delta _{jk} \right)^{2} }{4} } \end{array}\label{eq:dotProductsYang}\eeq 

The elements of the Gram matrix are given by (\ref{JonesOpTostokesInnerProd:eq})
\beq
\AA_{jk} =\sh_j \cdot \sh_k
= 2C_N^2 \left[ | \langle s_j |  s_k \rangle |^2 - \frac{1}{N}  \right]. 
\label{eq:dotproductsBasicEq}
\eeq 

Let's define the submatrices of Stokes vectors
\beq
\begin{array}{l} {X=\left[\hat{x}_{1} ,\ldots ,\hat{x}_{ N-1} \right]^{T} } \\[10pt] {Y=\left[\hat{y}_{1} ,\ldots ,\hat{y}_{\frac{N\left( N-1\right)}{2} } \right]^{T} } \\[10pt] {Z=\left[\hat{z}_{1} ,\ldots ,\hat{z}_{\frac{N\left( N-1\right)}{2} } \right]^{T} } \end{array}
\label{eq:YangBlocks}
\eeq 
where we reindexed the Stokes vectors from $\hat{y}_{ij} ,\hat{z}_{ij} $ to $\hat{y}_{k} ,\hat{z}_{k} $.

\begin{figure}[!ht]
	\centering
		\includegraphics[width=0.6\linewidth]{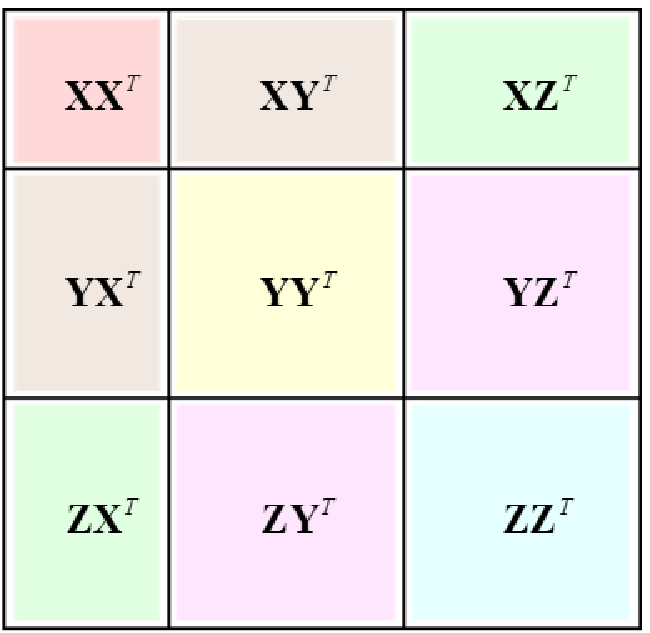}
		\caption{Partitioning of the Gram matrix for Yang and Nolan's vectors \cite{Yang}. Rectangular submatrices are shown in brown and green, while all other colors indicate square submatrices of different dimensions.}
	\label{fig:MatrixYang}
\end{figure}

The Gram matrix structure is shown in Fig. \ref{fig:MatrixYang}. The Gram matrix is partitioned into nine blocks (in color). An analytical calculation of the cost function in a way analogous to the case of SIC-POVMs and MUBs is cumbersome due to the more complex block structure of the Gram matrix. Therefore, the cost function shown in blue in Fig. \ref{fig:Ng1} is calculated numerically by taking the trace of the inverted Gram matrix obtained by (\ref{eq:dotProductsYang})-(\ref{eq:YangBlocks}).

\section*{Acknowledgment}
I. Roudas and D. A. Nolan would like to thank T. A. Nguyen, W. A. Wood, and J. Yang   of Corning Research and Development Corporation  for fruitful discussions. The authors would also like to thank the anonymous reviewers for their comments and suggestions that improved the quality of the manuscript.

% Can use something like this to put references on a page
% by themselves when using endfloat and the captionsoff option.
\ifCLASSOPTIONcaptionsoff
  \newpage
\fi

% trigger a \newpage just before the given reference
% number - used to balance the columns on the last page
% adjust value as needed - may need to be readjusted if
% the document is modified later
%\IEEEtriggeratref{8}
% The "triggered" command can be changed if desired:
%\IEEEtriggercmd{\enlargethispage{-5in}}

% references section

% can use a bibliography generated by BibTeX as a .bbl file
% BibTeX documentation can be easily obtained at:
% http://mirror.ctan.org/biblio/bibtex/contrib/doc/
% The IEEEtran BibTeX style support page is at:
% http://www.michaelshell.org/tex/ieeetran/bibtex/
%\bibliographystyle{IEEEtran}
% argument is your BibTeX string definitions and bibliography database(s)
%\bibliography{IEEEabrv,../bib/paper}
%
% <OR> manually copy in the resultant .bbl file
% set second argument of \begin to the number of references
% (used to reserve space for the reference number labels box)

\begin{IEEEbiographynophoto}{Ioannis Roudas}
	received his B.S. in Physics and an M.S. in Electronics and Radio-engineering from the University of Athens, Greece in 1988 and 1990, respectively, and an M.S. and a Ph.D. degree in coherent optical communication systems from the Ecole Nationale Sup\'{e}rieure des T\'{e}l\'{e}communications (currently T\'{e}l\'{e}com ParisTech), Paris, France in 1991 and 1995, respectively.
	
	During 1995-1998, he worked in the Optical Networking Research Department at Bell Communications Research (Bellcore), Red Bank, NJ. At the same time, he taught for two semesters,  as an Adjunct Professor, at Columbia University. He was subsequently with the Photonic Modeling and Process Engineering Department at Corning Inc., Somerset, NJ, from 1999 to 2002. During 2003-2011, he worked  as an Associate Professor of Optical Communications at the Department of Electrical and Computer Engineering at the University of Patras. In addition, he taught, as an Adjunct Professor, at the City University of New York and the Hellenic Open University. During 2011-2016, he was a Research Associate with the Science and Technology Division of Corning, Inc., Corning, NY. Since July 2016, he has been with the Department of Electrical and Computer Engineering at Montana State University as the Gilhousen Telecommunications Chair Professor.
	
	He is the author or co-author of more than 100 papers in scientific journals and international conferences and holds five patents. He currently serves as an Associate Editor for the IEEE Photonics Journal.
\end{IEEEbiographynophoto}

\begin{IEEEbiographynophoto}{Jaroslaw (Jarek) Kwapisz}
is a Polish-American mathematician with background in theoretical dynamical systems. He received M.S. (1991) degree from University of Warsaw and Ph.D. (1995) from State University of New York at Stony Brook. Jarek has worked on problems in several subject areas, including integral and differential equations, iterated maps modeling coupled non-linear oscillators, pattern formation in fourth-order Hamiltonian systems, ergodic theory and entropy in smooth and symbolic dynamics, cohomological Conley index and cocyclic subshifts, almost-periodic tiling spaces and quasi-crystals, abelian-Nielsen classes and geometry of translation surfaces, and conformal dimension of fractal sets. He is currently interested in quasi-symmetric renormalization for infinitely ramified fractals,  Anosov maps on infra-nil manifolds, non-Meyer substitution Delone sets, and applications of geometry of Stokes space to multi mode fiber-optic communication.
\end{IEEEbiographynophoto}

\begin{IEEEbiographynophoto}{Daniel A. Nolan}  received his Ph.D. degree from Pennsylvania State University in Physics in 1974. He currently is a
Corporate Research Fellow at Corning Research and Development Corporation. His research activities have
included the propagation of light in single- and multimode fibers, polarization optics, quantum optics, nonlinear effects in fibers, fiber-optic
sensors, and components for local area networks. He holds 101 U.S. patents.
He is also a Fellow of the Optical Society of America, and a recipient of
a number of awards including The IR100 award for the invention of infrared
polarizing glass, Polarcor\texttrademark, 1986; Corning’s “Outstanding Publication”
award in 1989; Corning’s Stookey award for Exploratory Research in 1995; the Penn State outstanding alumni award from the College of Science
for contributions to optical communications in 2001; and The Journal of Lighwave Technology Recognition award, 2008.
\end{IEEEbiographynophoto}

% biography section
% 
% If you have an EPS/PDF photo (graphicx package needed) extra braces are
% needed around the contents of the optional argument to biography to prevent
% the LaTeX parser from getting confused when it sees the complicated
% \includegraphics command within an optional argument. (You could create
% your own custom macro containing the \includegraphics command to make things
% simpler here.)
%\begin{IEEEbiography}[{\includegraphics[width=1in,height=1.25in,clip,keepaspectratio]{mshell}}]{Michael Shell}
% or if you just want to reserve a space for a photo:

%\begin{IEEEbiography}{Michael Shell}
%Biography text here.
%\end{IEEEbiography}

% if you will not have a photo at all:
%\begin{IEEEbiographynophoto}{John Doe}
%Biography text here.
%\end{IEEEbiographynophoto}

% insert where needed to balance the two columns on the last page with
% biographies
%\newpage

%\begin{IEEEbiographynophoto}{Jane Doe}
%Biography text here.
%\end{IEEEbiographynophoto}

% You can push biographies down or up by placing
% a \vfill before or after them. The appropriate
% use of \vfill depends on what kind of text is
% on the last page and whether or not the columns
% are being equalized.

%\vfill

% Can be used to pull up biographies so that the bottom of the last one
% is flush with the other column.
%\enlargethispage{-5in}

% that's all folks
\end{document}